\documentclass[journal,twoside]{IEEEtran}

\usepackage{setspace}
\usepackage{times,graphicx, amsfonts}
\usepackage[cmex10]{amsmath}
\usepackage{algorithmic}
\usepackage{array}
\usepackage{mathrsfs}
\usepackage{algorithm}
\usepackage{algorithmic}
\usepackage{epsf}
\usepackage[tight,footnotesize]{subfigure}
\usepackage{url}
\usepackage{amssymb}
\usepackage{cite}
\usepackage{multirow}
\usepackage{graphicx}
\usepackage{float}
\usepackage{booktabs}
\usepackage{indentfirst} 
\usepackage{xcolor} 
\usepackage{pifont}
\usepackage{xcolor,cite,etoolbox}
\makeatletter 
\pretocmd\@bibitem{\color{black}\csname keycolor#1\endcsname}{}{\fail}
\newcommand\citecolor[1]{\@namedef{keycolor#1}{\color{blue}}}
\makeatother

\begin{document}

\title{Energy and Information Management of Electric Vehicular Network: A Survey}

\author{Nan~Chen,~\IEEEmembership{Member,~IEEE,}
 Miao~Wang,~\IEEEmembership{Member,~IEEE,} 
 Ning~Zhang,~\IEEEmembership{Senior~Member,~IEEE,} 
        and~Xuemin~(Sherman)~Shen,~\IEEEmembership{Fellow,~IEEE} }

\maketitle

\begin{abstract}
The connected vehicle paradigm empowers vehicles with the capability to communicate with neighboring vehicles and infrastructure, shifting the role of vehicles from a transportation tool to an intelligent service platform. Meanwhile, the transportation electrification pushes forward the electric vehicle (EV) commercialization to reduce the greenhouse gas emission by petroleum combustion. The unstoppable trends of connected vehicle and EVs transform the traditional vehicular system to an electric vehicular network (EVN), a clean, mobile, and safe system. However, due to the mobility and heterogeneity of the EVN, improper management of the network could result in charging overload and data congestion. Thus, energy and information management of the EVN should be carefully studied.
In this paper, we provide a comprehensive survey on the deployment and management of EVN considering all three aspects of energy flow, data communication, and computation. We first introduce the management framework of EVN. Then, research works on the EV aggregator (AG) deployment are reviewed to provide energy and information infrastructure for the EVN. Based on the deployed AGs, we present the research work review on EV  scheduling that includes both charging and vehicle-to-grid (V2G) scheduling. Moreover, related works on information communication and computing are surveyed under each scenario. Finally, we discuss open research issues in the EVN.\let\thefootnote\relax\footnote{Corresponding author: Ning Zhang (ning.zhang@ieee.org)}\let\thefootnote\relax\footnote{N. Chen and X. Shen are with the Department of Electrical and Computer Engineering, University of Waterloo, Waterloo, ON N2L 3G1 Canada (e-mail: n37chen,  sshen@uwaterloo.ca).} \let\thefootnote\relax\footnote{M. Wang is with the Department of Electrical and Computer Engineering, Miami University, Oxford, OH 45056 United States (e-mail: wangm64@miamioh.edu).} \let\thefootnote\relax\footnote{N. Zhang is with the Department of Computer Science, Texas A$\&$M University at Corpus Christi, TX 78412, USA. (e-mail: ning.zhang@ieee.org).}
\end{abstract}

\begin{IEEEkeywords}
Electric vehicular network, connected vehicle, electric vehicle, energy scheduling, communication, computing.
\end{IEEEkeywords}

\section{Introduction}

The advancement of on-board sensing and communication technologies provides on-road vehicles with the capability to connect with surrounding vehicles and infrastructure, which is known as the connected vehicle paradigm \cite{sagning}. The empowered connectivity shifts the vehicle's role from a transport tool to an intelligent platform that can provide a wide range of services \cite{ITS2018}. For example, inter-vehicle connectivity provides vehicles with non-line-of-sight information to inform danger ahead, improve transportation safety \cite{auto3}. Meanwhile, infotainment service (\emph{e.g.}, online gaming) enhances the travelling experience while navigation service helps optimize the traffic condition \cite{sagning}. With its great potential, connected vehicle development has been promoted by academia, governments, and industries. Recently, dedicated short-range communication (DSRC) and cellular networks have been intensively developed to enable vehicle-to-everything (V2X) communication \cite{sagsurvey, 5gsurvey1}, supported by global automakers such as Honda, Nissan, and Toyota \cite{news}.

Another inevitable transition of the automobile system is transportation electrification due to the increasing environment concerns by petroleum combustion \cite{EPRI}. Equipped with large capacity electric batteries, electric vehicles (EVs) can utilize the local energy (\emph{e.g.}, hydro, wind, etc.) to charge the battery. By shifting the energy resource from petroleum to renewable energy, the greenhouse gas emission in the transportation sector can be effectively reduced. Moreover, on-board electric batteries provide EVs with energy storage potentials to provide vehicle-to-grid (V2G) service for enhancing the smart grid (SG) reliability \cite{V2GFR}. With the environmental advantages that EVs bring up, legislation has been launched worldwide to push forward the EV commercialization \cite{compgov}. For example, British Columbia has an incentive program which rewards \$2,500-\$6,000 for EV purchase or lease \cite{PEVincentivebc}. Moreover, the automobile industry is propelling the EV commercialization with grand EV manufacture plans. Besides EV-specialized company (\emph{e.g.}, Tesla), conventional car industry such as BMW plans to enable EV manufacture on most of their vehicle models by 2021 \cite{bmw}.

The unstoppable trends of the connected vehicle and transportation electrification transform the traditional vehicular system to an electric vehicular network (EVN), where EVs are connected on-the-move to perform intelligent service for the SG and transportation network (TN). As the number of EVs and their aggregators (AGs) continues growing, they can be used as additional communication/computing infrastructure to facilitate the EVN operation. Correspondingly, the increasing connectivity of the  EVN provides EVs with the latest update from the SG and TN. In the case of emergency, connected EVs can be directed to provide a variety of ancillary services. The environmental and technical advantages of the EVN motivate a transition towards a mobility system that is clean, connected, and safe, which is also aligned with the TN objective of major countries and unions \cite{auto3}.

\begin{figure*}[!t]
\centering
\includegraphics[width=0.8\textwidth]{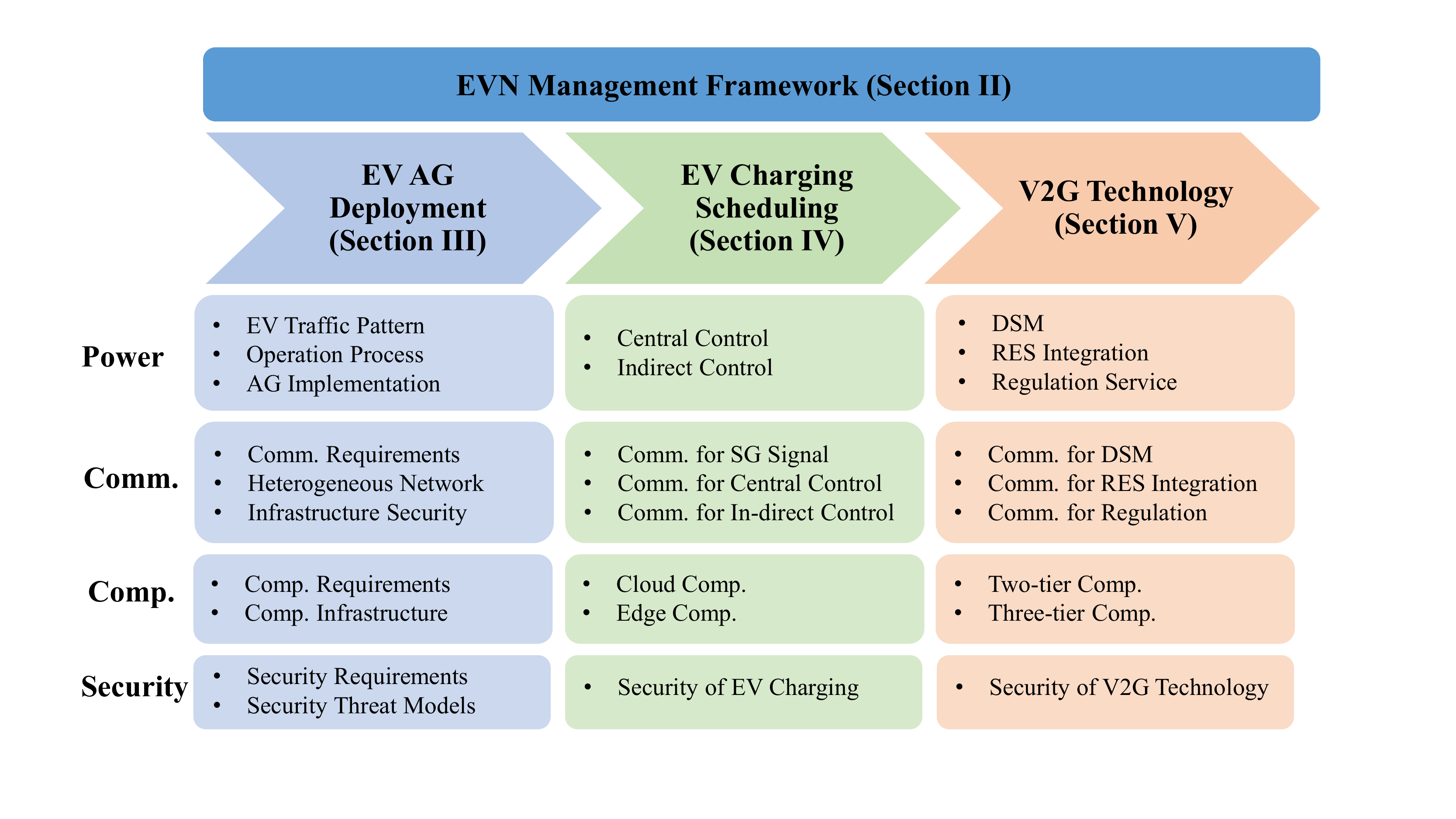}
\vspace{-0.3cm}
\caption{Paper organization.}
\vspace{-0.3cm}
\label{Fig_con}
\end{figure*}

Due to the variety of technologies and networks that the EVN adopts, many technical challenges arise.
From the energy management perspective, as the interface between EVs and the SG, AGs encounter deployment issues caused by EV mobility, the heterogeneity of charging standards, and other service concerns. Moreover, the EV charging overloading impact on the SG is critical to system stability. If the market share of EVs accounts for $62\%$ in the automobile market as predicted in \cite{EPRI}, simultaneous charging of multiple EVs at peak hours poses great challenges on the aged power infrastructure \cite{surveyelectric}. Besides, the V2G service not only requires a large number of EVs, but also exaggerates range anxiety and battery degradation concerns for EV drivers, impeding smooth V2G operation \cite{rangeanxiety}.
From the information management perspective, a connected vehicle is expected to generate on average 25GB/h per day, leading to a data burst that can be overwhelming for the traditional communication network. Although the fifth generation (5G) cellular network is expected to be a potential solution, mobile EVs still have limited connectivity in rural areas where the territorial network cannot fully cover. A promising method is to integrate space and aerial networks to the ground network, forming a widely covered and cost-efficient space-air-ground (SAG) network. However, the multi-dimensional network structure of SAG and the heterogeneity in devices (\emph{e.g.}, satellite, airship, EVs, etc.) demand an efficient resource allocation scheme to guarantee the network performance \cite{nansdn}.

Facing these challenges, researchers have put tremendous efforts to address research issues on EV energy and SG information management. In the literature, several surveys and tutorials focusing on the energy management of EVs in SG. 
As an outlook of the EV, \cite{infrasr} presents the current states of battery technology, EV charging standards, and EV charging infrastructures. With EV commercialization proceeding rapidly, the impacts of EV charging on the SG have also been extensively researched.
\cite{impactsurvey} studies the charging impact on the system component level, while \cite{impactr2015} investigates the impact broadly on the economy, environment, and system stability.
Based on the EV charging impact study, \cite{whplanning} reviews the charging infrastructure deployment from the SG perspective.
Both charging and discharging scheduling has been comprehensively reviewed from the viewpoints of methodology \cite{review2014, surveyelectric, LHsurvey, EV5} and algorithm \cite{smartchargings}. 
On the other hand, the communication technologies and infrastructure in the SG are extensively reviewed. The communication infrastructure of the SG is discussed in terms of design and implementation in \cite{2014sg, sgcomm2, 2017sg, wavecomm}. Comprehensive surveys on the SG communication requirements and security issues are presented in \cite{shsecurity, securitydata}. Surveys on the application of cloud/edge computing in the SG are presented in \cite{cloudSG, IoTFog}.

The existing surveys and reviews provide insights on either EV energy management or SG communication/computing technologies where EVs are only considered as a small component. Most of them are focused on isolated topics, while the discussion on inter-system operation remains open. As EVs become prevalent in the automobile market, they are placed in the unique positions that interconnect power, communication, and computing networks. On one hand, vehicular communication and computing technologies enhance the EV charging/discharging performance via ubiquitous communication coverage and fast computing. On the other hand, EVs have the potentials to be mobile data relays and computing nodes to facilitate on-road information transmission and computing. Nevertheless, the mobile properties of EVs could impede the management of such an interconnected network, as the mobility brings in a variety of new operating scenarios in the EVN that has stringent energy and information requirements. Therefore, a thorough review of the inter-system operation of the EVN is in urgent demand.

To the best of our knowledge, we are the first to present the concept and state-of-the-art review on the EVN management considering the interoperation of power, communication, and computing. Owing to the mobile and electric nature of EVs, our introduced management framework is composed of the SAG-integrated vehicular network for wide communication coverage, a hierarchical computing infrastructure for timely operation, a software-defined network (SDN) controller for flexible operation, and electric components. In terms of the EV commercialization process, existing literature of EVN energy/information management is reviewed by stages: from EV AG deployment to EV charging to V2G technology. As the crucial interfaces between EVs and the SG, AGs are expected to be regional operators that are equipped with communication and computing capabilities. In this paper, we first provide a detailed review of AG deployment in the SG, considering aspects of power, communication, and computing. Further, we conduct in-depth surveys on scheduling schemes of EV charging and V2G technology concerning power analysis, data communication, computation, and security. Finally, we outline open issues and future research topics for EVN management by stages.

A list of acronyms used throughout the paper is presented in Table \ref{acr}. The organization of the paper is shown in Fig. \ref{Fig_con}. The management framework of the EVN is presented in Section II. The state-of-the-art survey of AG deployment is provided in Section III, where the roles of EV AGs as energy interfaces, their related communication and computation infrastructure are discussed. Surveys on the latest research works of EV charging and V2G technology are presented in Sections IV and V, respectively. The challenges and related works of communication, computation, and security under each scheduling scenario are also carefully discussed. The open research issues under different EV commercialization stages are presented in Section VI. Finally, the conclusion is given in Section VII.

\begin{table}[!t]
\centering

\setlength{\abovecaptionskip}{0pt}
\footnotesize

\caption{LIST OF ACRONYMS AND DEFINITION}
\label{acr}

\renewcommand\arraystretch{1}
\begin{center}
\begin{tabular}{|l|l|}
	\hline
	\textbf{Acronyms} & \textbf{Definition} \\ 
	\hline
	5G & Fifth Generation \\
	\hline
	AC & Alternating Current \\
	\hline
	AG & Aggregator \\
	\hline
	BEV & Battery Electric Vehicle \\
	\hline
	DC & Direct Current \\
	\hline
	DoD & Depth-of-Discharge \\
	\hline
	DSM & Demand Side Management \\
	\hline
	DSRC & Dedicated Short Range Communication \\
	\hline
	ECU & Electrical Control Units \\
	\hline
	EV & Electric Vehicle \\
    \hline	
    EVN & Electric Vehicular Network \\
    \hline
    HAN & Home Area Network \\
    \hline
     IaaS & Infrastructure as a Service \\
    \hline    
    IEC & International Electrotechnical Commission \\
    \hline
    IoT & Internet-of-Things \\
    \hline
    IPT & Inductive Power Transfer \\
    \hline
    ISO & International Organization for Standardization \\
    \hline
    MCS & Monte Carlo Simulation \\
    \hline
    MILP & Mixed Integer Linear Programming \\
    \hline 
    MIMO & Multiple-Input Multiple-Output \\
    \hline
    NAN & Neighbourhood Area Network \\
    \hline
    OBU & On-Board Unit \\
    \hline
    PaaS & Platform as a Service \\
    \hline
    PHEV & Plug-in Hybrid Electric Vehicle \\
    \hline
    PLC & Power Line Communication \\
    \hline
    QoS & Quality-of-Service \\
    \hline
    RES-DG & Renewable Energy Source - Distributed Generation \\
    \hline
    RD/RU & Regulation Down/Regulation Up \\
    \hline
    RSU & Roadside Unit \\
    \hline
    SaaS & Software as a Service \\
    \hline
    SAE & Society of Automotive Engineers \\
    \hline    
    SAG & Space-Air-Ground \\
    \hline	
    SDN & Software-Defined Network \\
    \hline
    SE & Social Equilibrium \\
    \hline
	SG & Smart Grid \\
	\hline
	SoC & State-of-Charge\\
	\hline
	SVM & Support Vector Machine \\
	\hline
	TN & Transportation Network \\
	\hline
	ToU & Time-of-Use \\
	\hline
	UAV & Unmanned Aerial Vehicle \\
	\hline
	UE & User Equilibrium \\
	\hline
	URLLC & Ultra-Reliable Low-latency Communication \\
	\hline
	V2B & Vehicle-to-Building \\
	\hline
	V2H & Vehicle-to-Home \\
	\hline
	V2I & Vehicle-to-Infrastructure\\
	\hline
	V2G & Vehicle-to-Grid \\
	\hline
	V2X & Vehicle-to-Everything \\
	\hline
    V2V & Vehicle-to-Vehicle \\
    \hline
    VANETs & Vehicle ad hoc Networks \\
    \hline
    WAN & Wide Area Network \\
    \hline
    WLAN & Wireless Local Area Network \\
    \hline

\end{tabular}
\end{center}
\end{table}

\section{The Management Framework of EVN}

\begin{figure}[!t]
\centering
\includegraphics[width=0.5\textwidth]{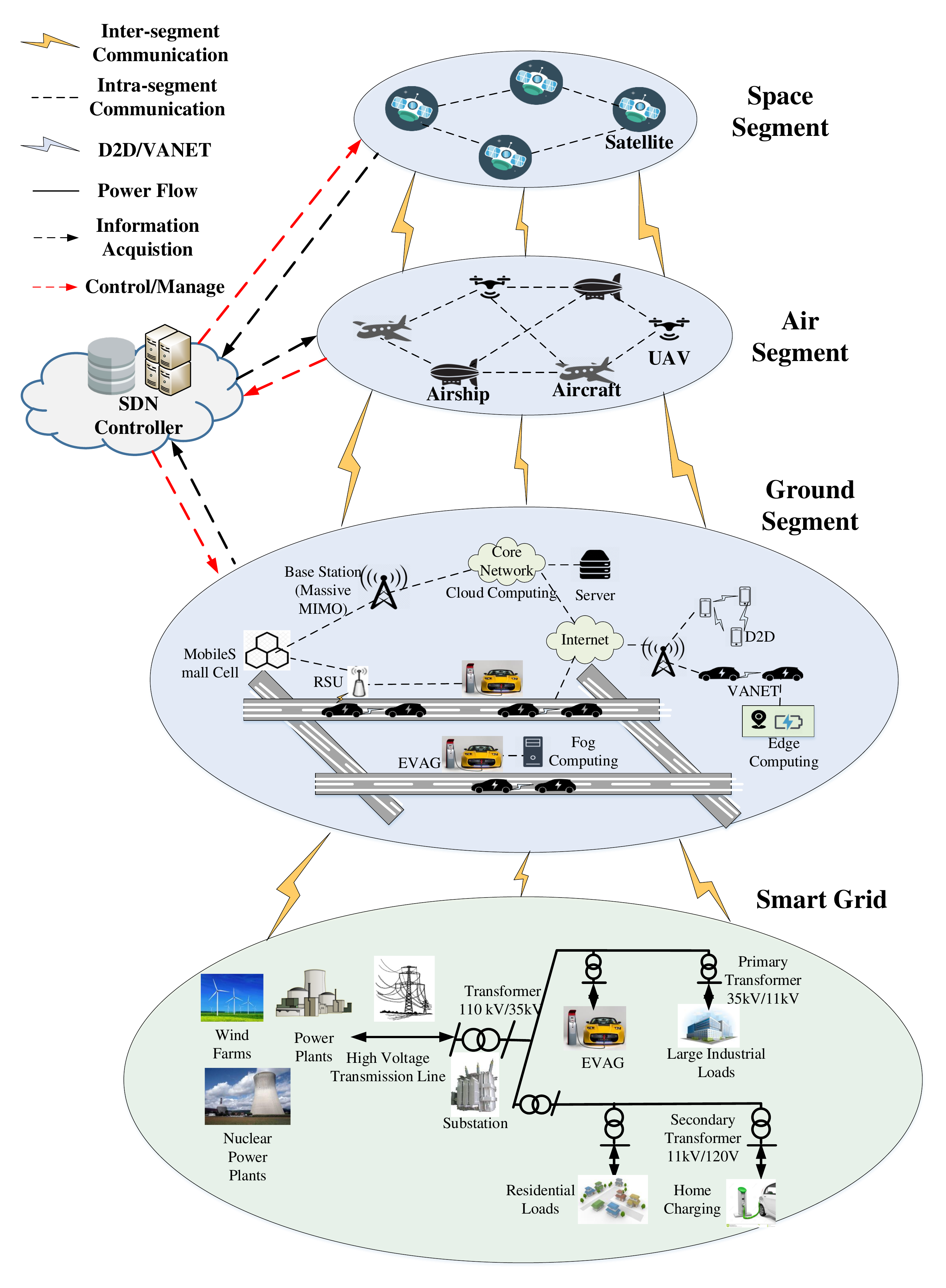}
\vspace{-0.3cm}
\caption{The management framework for the EVN.}
\vspace{-0.3cm}
\label{Fig_sag}
\end{figure}

The mobility and electricity demand of EVs require the management of communication, computation, and power systems \cite{luning}. Thus, a comprehensive management framework for EVN is required, as shown in Fig. \ref{Fig_sag}. The framework consists of the SAG-integrated vehicular network for efficient data transmission, computing infrastructure to distribute service tasks accordingly, and SDN control for dynamic, efficient system-level control. Moreover, as crucial components of the EVN, EVs and AGs are introduced in this section. Next, we introduce each part in details, finished by summarizing the section with existing challenges in the EVN.

\subsection{SAG-Integrated Vehicular Network}

To support the EV operation (\emph{e.g.}, charging, V2G, navigation, etc.) on-the-move, heterogeneous network frameworks are proposed in \cite{hetsurvey, TVTzhuang} to connect vehicles. Two mainly adopted communication techniques are dedicated short distance communication (DSRC) and cellular networks \cite{sagning, hetsurvey, TVTzhuang}. DSRC facilitates both vehicle-to-vehicle (V2V) and vehicle-to-infrastructure (V2I) communication while the cellular network provides reliable Internet access. However, DSRC requires a large-scale network infrastructure deployed to enable timely communication \cite{vanetintro}. The cellular network could encounter communication congestion in urban areas while the coverage at rural areas is very poor. Furthermore, both DSRC-based and cellular networks have difficulty in supporting highly mobile vehicles, due to the frequent handovers \cite{handover1}.

To address the coverage and handover issues, the communication operators are developing satellite and aerial networks to facilitate a multi-dimensional communication network, named as the SAG integrated vehicular network \cite{sagning}. Recently, an increasing number of projects are under way on both customer (vehicle) and supplier sides. For example, Toyota's Mirai Research Vehicle can provide a mobile communication service at the data rate of 50 Mbps with mTenna \cite{sagning}. Tesla and Google both plan to launch satellites for Internet access \cite{googleicc}. The satellite networks have been in the space for decades, mainly for navigation, earth observation, and other communication/relay services. With the advanced astronautic technology and virtualization techniques, satellite networks are expected to help the ground network offload communication tasks\cite{jsacoffload}. With the advantages of wide coverage, reliable access, and multi-casting capability, satellite networks can provide high data-rate coverage at rural areas to complement the coverage problem of cellular networks \cite{sagsurvey}.

Although the satellite network can provide great coverage at rural areas, its long propagation delay and limited flexibility cannot efficiently alleviate the heavy data communication tasks in urban areas \cite{jsaclatency}. In this case, the aerial network that is formed by unmanned aerial vehicles (UAVs), balloons and airships at the stratosphere can provide broadband connectivity with extended coverage. Moreover, the controllability of the network devices enhances the system flexibility by offloading communication tasks as directed \cite{flexibilitymag}.

The ground network at the bottom layer of the vehicular network consists of cellular network, DSRC, and so on. By adopting technologies such as massive multiple-input multiple-output (MIMO), millimetre wave, spectrum sharing, the fifth generation (5G) cellular network provides ultra-reliable and low-latency communication (URLLC) \cite{5gsurvey1, 5gsurvey2, 5gsurvey3}. In addition to cellular networks, depending on the vehicle status (\emph{i.e.}, mobile and stationary), other networks and technologies are adopted to satisfy a variety of EV services, which will be discussed in detail in Section III-E.

\subsection{Computing Infrastructure}

\begin{figure}[!t]
\centering
\includegraphics[width=0.5\textwidth]{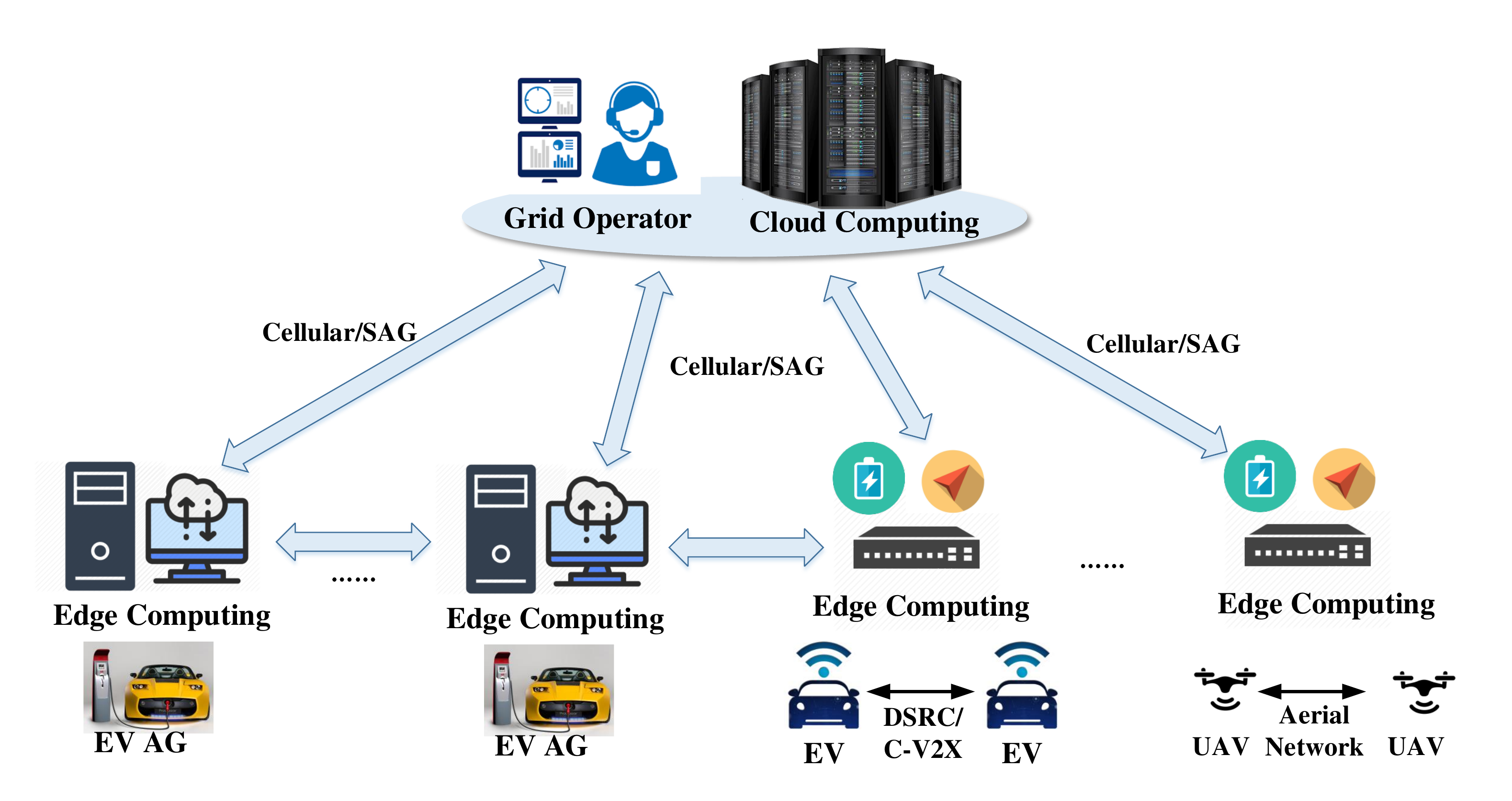}
\vspace{-0.3cm}
\caption{The computing architecture of the EVN.}
\vspace{-0.3cm}
\label{Fig_computing}
\end{figure}

As predicted by IHS automotive company, by 2020, there will be 152 million actively connected vehicles on the road, with an average 30 TB data produced per vehicle per day \cite{HCFog}. The huge data generation demands large-size data storage. Moreover, due to the heterogeneity in data context and computing, the complexity of data processing and analysis is extremely high. Hence, a hierarchical computing architecture is needed for the EVN to enhance the system data storage and computing capabilities \cite{fogmag}, as shown in Fig. \ref{Fig_computing}. It is composed of the remote cloud computing platform with adequate resources and edge computing with resources in proximity to EVs.

As an emerging computing model, cloud computing is a shared pool with configurable resources and services that can be easily accessed. With access to remote software and hardware, operators can process and analyze the data with low-cost \cite{cloudsurvey}. In the EVN, cloud computing is usually deployed at the SG level with large storage and computation device, as the brain of the SG. The computation centre constantly collects and stores operation data of EVs and AGs. Then, operation algorithms (\emph{e.g.}, demand response, economic dispatch, PEV charging/discharging management, etc.) are implemented on the cloud to provide the grid with the optimal operation \cite{cloudSG}. However, the enormous data volume generated by increasing Internet-of-Things (IoT) devices and connected vehicles poses great pressure on the cloud. The cloud can encounter severe computation delay and high communication latency. Thus, edge computing emerges as a novel paradigm to help alleviate the heavy computing tasks at the cloud and reduce service latency for time-critical applications in EVN.

Edge computing provides computation and storage resources in proximity to end devices to accomplish local computing tasks \cite{fog1}.  The geographic distance between edge nodes and end users are closer, and therefore, communication overhead can be effectively reduced. By undertaking local computing tasks, edge nodes not only reduce the task response time, but also alleviate the data storage burden at the cloud \cite{HCFog}. For example, EV AG can collect neighbouring traffic and EV information, and perform charging/discharging scheduling locally using in-station computing devices. Further, a group of vehicles can be considered as opportunistic edge computing nodes that collect neighbouring traffic and vehicular information to either process the data in-vehicle or transmit the data to nearby AGs. The dense coverage, high computation capacity, and decentralized management of edge computing provide low-latency, high-security, and high-quality computing service \cite{edge1}.

The introduced computing architecture enables a low-latency, high-computation, and secure EVN operation. As the IoT devices and vehicles increase in the EVN, the management of the computing architecture needs to be scalable, flexible, and secure. In this case, the emerging SDN control can be a promising option for energy/information management in the EVN\cite{edgesurvey}.

\subsection{SDN Control}

\begin{figure}[!t]
\centering
\includegraphics[width=0.4\textwidth]{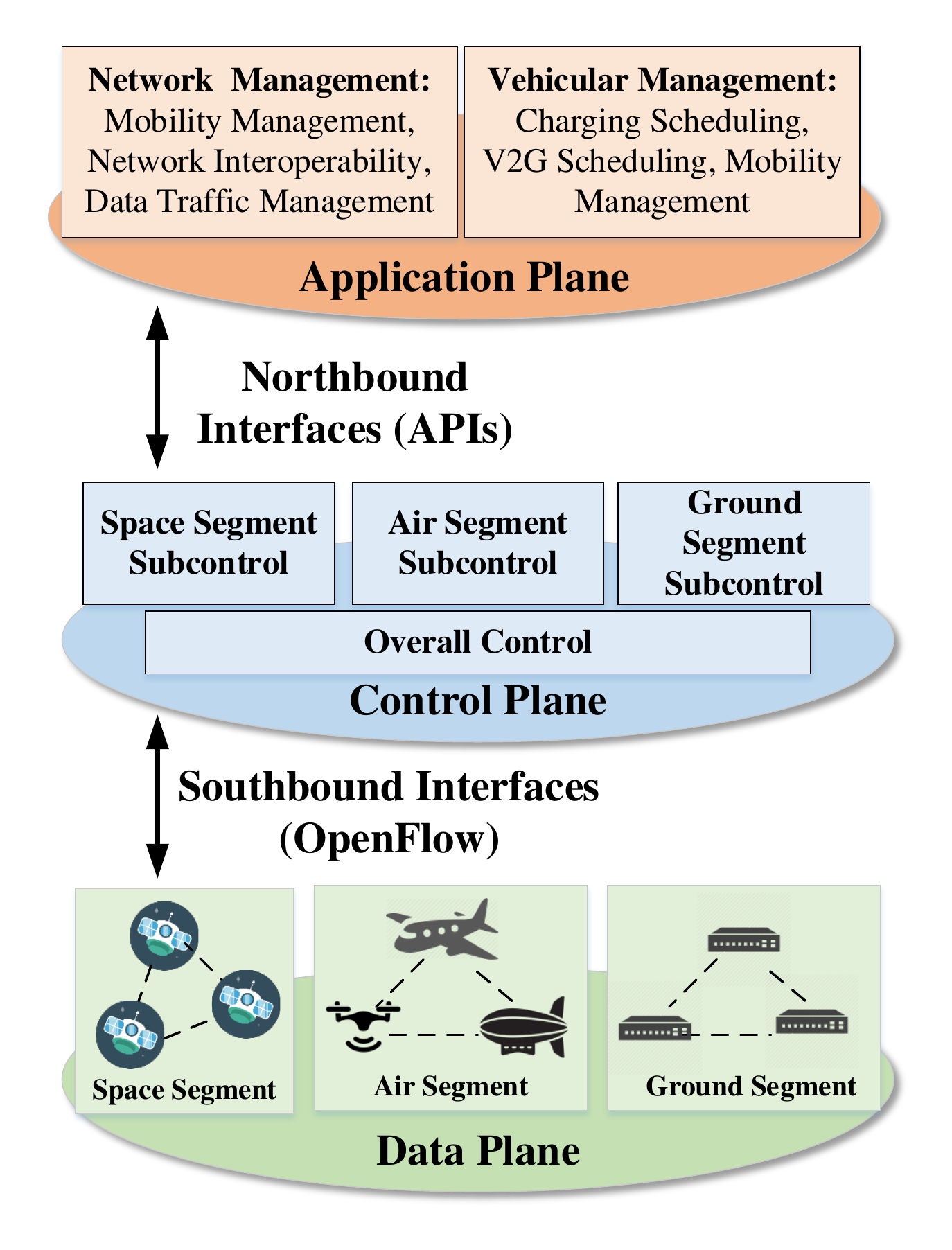}
\vspace{-0.3cm}
\caption{The architecture of SDN control.}
\vspace{-0.3cm}
\label{Fig_sdn}
\end{figure}

To effectively operate the EVN that intersects power, communication, and computing systems, there are many challenges:

\begin{itemize}

\item \textbf{SG Operation}: As more renewable energy and EVs are integrated to the SG, the power profile of the SG fluctuates spatially and temporally. To enable energy balance at all-time in the SG, the grid operator demands automatic and intelligent management schemes \cite{nansdn};

\item \textbf{Communication}: Considering the diversity in the SG IoT devices, conventional communication methods cannot guarantee a flexible data transfer among devices\cite{SDNsurvey2019}. For example, when a new service is added in the SG, every router needs to be reconfigured, which results in high manual cost and service disruption. The situation could be more complicated when integrating SAG networks to the EVN. Challenges arise such as inter-operation of satellite, aerial and territorial communication technologies, (dynamic) network management, and Quality-of-Service (QoS) provision \cite{sagning}.

\item \textbf{Computing}: The complex structure of the EVN requires a hierarchical computing architecture to ensure that data stored/analyzed at different layers can meet computing demand correspondingly \cite{EVTVT}. The management of the computing architecture in terms of data collection, storage, and computing requires a flexible and automated operation mechanism \cite{edge1}.

\end{itemize}

The emerging paradigm, SDN, has great potentials to be a promising management method for the EVN to address the above challenges As shown in Fig. \ref{Fig_sdn}, SDN decouples the control intelligence from physical devices (\emph{e.g.}, switches) and forms a control plane to control devices centrally. With its programmability at the application planes and open interfaces between control and data planes, SDN enables a dynamic, flexible interoperable, and cost-efficient network.

By applying SDN to the power operation, the grid operator can perform a variety of services (\emph{e.g.,} EV charging, V2G, demand response, etc.) on the application plane. Under different operating scenarios, the grid operators can quickly perform the service analysis and distributes command to the physical devices on the data plane, enabling an intelligent and fast-response operation performance \cite{nansdn}.

In the communication system, by extracting intelligence from physical devices and configuring data forwarding rules with the control plane, SDN makes the data communication cost-efficient and flexible \cite{sagning}. The open interface feature also enables the interoperability among a variety of communication technologies. Further, in the SAG-integrated vehicular network, the control plane is partitioned into different control segments targeted at network operation in different domains to suit individual communication features \cite{sagning}.

The management of computing infrastructure becomes more flexible and efficient supported by SDN. \cite{EVTVT} proposes a hierarchical SDN management framework to handle specific tasks and issues on different tiers. Another approach is to implement computing tasks through virtualization technology to partition the computing resource into various slides for different services \cite{yeqiang}. For example, the satellite network can be partitioned into different slices. While traditional tasks (\emph{e.g.}, navigation, earth observatory) take up several slices as demanded, the rest available slices can help support the ground network and other functions \cite{mcsatellite}.

\subsection{EV}

According to \cite{def2}, EVs are defined as the vehicles that can be charged from an external electric source through plugs or wall sockets. Electricity is stored in on-board rechargeable batteries for mechanical propulsion. In terms of propulsion sources, EVs can be further categorized into two groups: plug-in hybrid electric vehicles (PHEVs) and battery electric vehicles (BEVs). PHEVs are equipped with both electric motor and internal combustion engine while BEVs are pure EVs with large battery capacities. Currently, manufactured BEV has an average driving range of 150 miles with a 40 kWh battery capacity \cite{nissan}. Advanced BEVs such as the Tesla Model S with 85kWh battery can travel up to 335 miles with one single charge \cite{teslas}.

EVs are communication-enabled with on-board units (OBUs) to help them connect with other vehicles and infrastructure among the EVN. Bluetooth is a commonly adopted technology for inter-vehicle connection, while DSRC is gradually integrated among EVs for V2V and V2I data transfer \cite{vanet1}. Satellite communication is also enabled to connect the vehicle in rural areas. Within the vehicle, more and more sensors are deployed for tire pressure monitor, temperature detection, location detection, etc. Intra-EV communication technology (such as controller area network, time-triggered Ethernet, etc.) is applied among sensors and electrical control units (ECU) \cite{luning}. ECU analyzes the sensed data to adjust EV mechanical performance and battery management for the optimal EV operation. As more and more EVs travel on-road, they can be deemed as edge computing nodes to sense and communicate with the surrounding environment and conduct local computing tasks \cite{edge1}.

\subsection{EV AG}

EV AG is the EV integration facility that can charge/discharge multiple EVs simultaneously. private EV chargers and public EV charging stations are two main types of AGs. In terms of different charging demands, EV AGs are deployed at feeders with different voltage/power capacities, as shown in Fig. \ref{Fig_sag}. Public charging stations are deployed at primary feeders while private EV charging points are deployed at secondary feeders \cite{nanTVT}.

\begin{figure}[!t]
\centering
\includegraphics[width=0.5\textwidth]{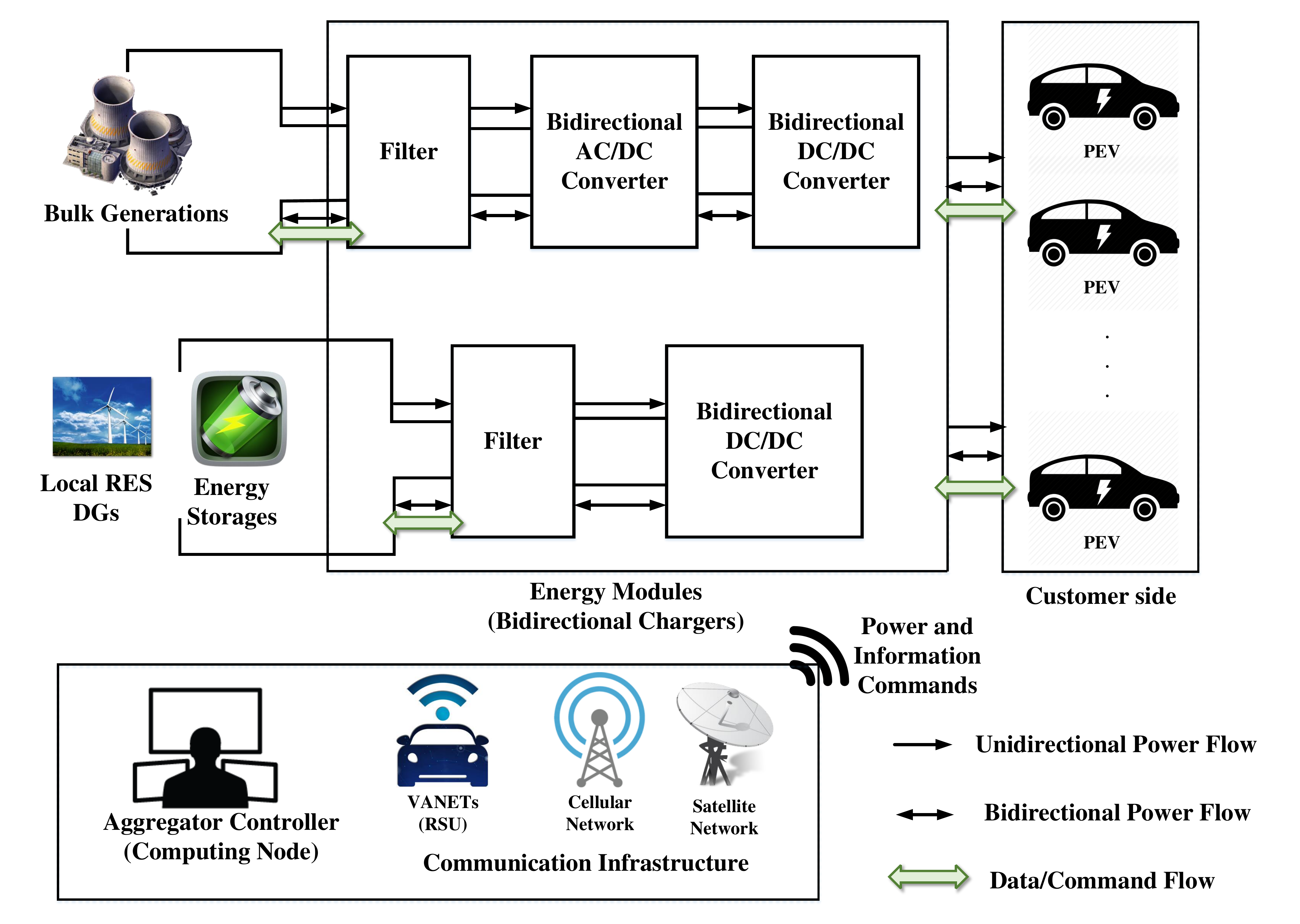}
\vspace{-0.3cm}
\caption{Structure of EV charging AG.}
\vspace{-0.3cm}
\label{Fig_charging_infra}
\end{figure}

The basic structure of the AG consists of controller, energy modules (bidirectional chargers), and communication/computing infrastructure, while bulk generations and renewable energy source-distributed generators (RES-DGs) are connected to AG for power supplement, as shown in Fig. \ref{Fig_charging_infra}.

\subsubsection{Bidirectional Chargers}

The energy modules (\emph{i.e.}, bidirectional chargers) have two types of modes with respect to different power sources. When connected to bulk generations, the module first removes the unwanted frequency by filters. The alternating current (AC)/direct current (DC) converter enforces power factor while the DC/DC converter regulates the battery current \cite{infrasr}. When charging EVs, the chargers draw AC from the bulk generations with a defined phase angle. When discharging EVs, the chargers return the same form of current back to the SG.
RES-DGs that use local source are also a supply option \cite{nanwcsp}. Considering the DC nature of most RES-DG \cite{dcres}, DC/DC converter is used to regulate the battery charging current at a desired level. 

To facilitate the interoperability between EV AGs and EVs, charging standards are proposed worldwide and categorized in AC and DC level, as summarized in Table \ref{chargingstandard}. AC charging takes up to 17 hours to fully charge a BEV \cite{SAEJ1772}, regarding as slow charging.  AC charging is usually adopted at private chargers for overnight charging. DC charging has a faster speed with the charging time ranging from minutes to 1.7 hours \cite{compare}. DC charging is mainly adopted at public charging stations for EVs on the move.

\begin{table}[!t]
\centering

\setlength{\abovecaptionskip}{0pt}
\scriptsize

\caption{CHARGING STANDARD}
\label{chargingstandard}

\renewcommand\arraystretch{1.5}
\begin{center}
\begin{tabular}{|p{70pt}|p{80pt}|p{50pt}|}
	\hline
	Charging Method & Charging Power Requirement & Estimated Charging Time\\ 
	\hline
	AC level 1 & 120 VAC, 12 A, 1-phase, 1.4-1.9 kW & 7-17 hours\\
	\hline
	AC level 2 & 208-240 VAC, $\leqslant$ 80A, 1/3-phase, $\leqslant$ 19.2 kW & 22 min-3 hours \\
	\hline
	SAE CCS (DC) & 200-600 VDC, 80-400 A, $\leqslant$ 240 kW & $\leqslant$ 1.2 hours \\
	\hline
	CHAdeMO (DC) & 417 VDC, 120 A, 50 kW & $\leqslant$ 1.7 hours \\
	\hline
	SuperCharger (DC) & 450-600 VDC, 200-225 A, 90-120 kW & $\leqslant$ 1 hour \\
	\hline
	
\end{tabular}
\end{center}
\end{table}

\subsubsection{AG Controller}

The AG controller is responsible for scheduling nearby EV services (\emph{i.e.}, charging, V2G, navigation, and communication) as a fog node in the introduced framework where the AG communicates with neighbouring EVs for data collection and analysis. While some tasks can be accomplished locally (\emph{e.g.}, charging), other tasks (V2G, navigation) submit the processed date back to the cloud node for an overall system review.
 
\subsubsection{Data Communication}

As the backbone for EV scheduling, data communication is essential to both EV AG and EVs. When EVs move on-road, their OBUs exchange vehicle data (\emph{e.g.}, State-of-Charge (SoC), charging requests, etc.) with AG controller. When EVs are stopped in-AG, communication is also needed to negotiate charging/discharging processes \cite{surveyelectric}. Moreover, AGs also exchange their operation conditions with the SG to synchronize AG information. Under a variety of services and application scenarios, different technologies and networks are demanded in the vehicular network, forming a heterogeneous network \cite{2013comm}.

To improve the communication interoperability, organizations around the world such as International Organization for Standardization (ISO), Society of Automotive Engineers (SAE), and International Electrotechnical Commission (IEC) have developed standard for SG communication such as ISO 15118, J2847, IEC 61850,etc \cite{2015standard}.

\subsection{Management Challenges}

The mobility and electrification of vehicles introduce a large number of EVs as mobile and high power rating appliances to the SG and vehicular network, which poses many challenges to the EVN management:

\begin{itemize}
\item \textbf{EV AG Deployment} - The deployment of EV AG not only determines the AG service provision profit, but also affects EV driving range. Moreover, the implementation of information infrastructure is closely related to the AG QoS, hence, complicating the AG deployment;

\item \textbf{EV Charging Scheduling} - Simultaneous charging of a large number of EVs could jeopardize the aged power infrastructure, while the data transfer among mobile EVs could encounter frequent handovers and non-negligible latency;

\item \textbf{V2G scheduling} - While V2G can bring additional power supports for the SG and economic benefits for EV drivers, the V2G scheduling still faces challenges considering the large amount of EVs demanded for ancillary service. Moreover, time-sensitive V2G service requires seamless communication, which sets a high standard for the vehicular network.

\end{itemize}

To overcome the challenges, extensive research has been conducted by managing both energy and information aspects of the EVN. Next, we review the existing research works in detail to address the above challenges, respectively.

\section{EV AG Deployment}

Charging EVs in the power distribution system incurs many challenges due to EV mobility and their heavy electricity demands. Simulation in \cite{impact2010} shows that a 30$\%$ EV integration to the residential grid can incur significant voltage deviation. Extensive integration of EVs can also increase the overhead distribution transformer aging factor \cite{impact2015} due to excessive operating temperature. 
To mitigate EV charging impact on the SG, it is essential for the grid operator to determine the service capacities of AGs in terms of the SG constraints and charging demands. 
On the other hand, as the interface between EVs and the SG, AG undertakes the communication tasks between both sides. While the SG needs reliable and timely data communication with AG, EVs demands mobile communication. Thus, the communication infrastructure for AG is also essential at the deployment stage. Finally, the computing infrastructure upgrade at AG is introduced to enhance the whole system computing efficiency.

Next,  we review the works related to the EV AG deployment from both energy and information aspects by first identifying the challenges. Then,  related works on the AG deployment will be discussed in detail step by step in the remainder of the section.

\subsection{Deployment Challenges}

The public AGs are deployed either along the road, or in the parking lots in commercial and industrial areas. Compared with private AGs, the installation cost per charger at the public AGs is lower, since aggregated EVs are charged through the same transformer instead of connecting EVs to different transformer as in the private case \cite{coneco1}. Considering the limited controllability of private AGs, we mainly discuss the public AG deployment in this section, which has its unique challenges:

\begin{itemize}
\item EV mobility leads to highly fluctuating charging demands, increasing the estimation error of AG service capacities;

\item In terms of different charging standards and vehicle modes, AG operation modelling also has its stochastic properties against the AG service capacity estimation;

\item As a profitable investment, AG takes numerous economic factors (\emph{e.g.}, installation cost, service time, traffic intensity, power system constraints, etc.) into consideration, making the deployment complicated;

\item The extensive communication range incurred by vehicular mobility demands a well-designed communication infrastructure;

\item The foreseen large computation tasks and various computation requirements incurred by different EV service tasks demand a hierarchical computing architecture.

\end{itemize}

Research works on addressing the energy-related deployment challenges are first reviewed. Then, the information management related work is introduced to address the last two challenge.

\subsection{EV Travel Pattern Analysis}

EV mobility brings unpredictable errors to the demand forecast of the EV AG. Hence, to estimate the charging capacity of the AG, EV travel pattern analysis is essential as the first deployment step. As the AG deployment is conducted on a long time-line, the traffic pattern analysis results are macroscopic traffic data (\emph{e.g.}, traffic flow, density, speed, etc.). Although numerous travel pattern research works have been published in recent decades, there only exists a small portion of EV-related travel pattern works due to the lack of EV travelling data \cite{whplanning}. In this section, the EV-related travel pattern works are reviewed in terms of their modelling methodologies, as shown in Table \ref{traffic}.

\begin{table}[!t]
\centering

\setlength{\abovecaptionskip}{0pt}
\scriptsize

\caption{Modelling Methodology of EV traffic pattern}
\label{traffic}

\renewcommand\arraystretch{1.5}
\begin{center}
\begin{tabular}{c|c}
	\hline
	Model Category & Detailed Model \\ 
	\hline
	
	\multirow{3}{*}{\centering{Data-based model}} & Survey source \cite{usnational, britishdata, pems, 2017online, 2015canada} \\
	\cline{2-2} & Trace-based model \cite{mobility2009, beijing} \\
	\cline{2-2} & Data mining \cite{datamag, datamag2, SVM} \\
     \hline
	\multirow{4}{*}{\centering{Synthetic model}} & MCS \cite{mostafamodel} \\
	\cline{2-2} & Queueing model \cite{BCMPM} \\
	\cline{2-2} & Fluid dynamic traffic \cite{spatial} \\
	\cline{2-2} & Traffic assignment \cite{MCS2016, somodel, stochastic2018}\\
	\hline
	Simulator-Based Model & Commercial software \cite{miaomag, VISSIM, microscope, SUMO, PARAMICS, CORSIM} \\
	\hline
	
\end{tabular}
\end{center}
\end{table}

\subsubsection{Data-Based Model}

For the data-based model, the sources of data are essential as they are the basis of a realistic and precise travelling model. The U.S. Federal Highway Administration regularly conducts national household travel survey that includes daily non-commercial travel data by all vehicle modes \cite{usnational}. British national travel survey also reports statistics that cover personal travel \cite{britishdata}. The state of California also collects real-time data of California freeway, which can be obtained from their PeMS server \cite{pems}. Recently, more EV-related data have been surveyed to examine the travelling trends and patterns of EV users. For instance, \cite{2017online, 2015canada} survey the EV usage, daily travelling amount, fuelling condition, and predicted results.

\textbf{Trace-based Model:} By directly extracting generic travelling pattern from the data, the model provides time-efficient and effective results \cite{mobility2009, beijing}. However, this modelling scheme heavily depends on the quantity, quality, and vehicle property of the extracted data. For example, the taxi travelling trajectory extracted from the Beijing taxi data \cite{beijing} cannot represent personal vehicle usage.

\textbf{Data Mining:} As a process to discover pattern in large data sets, data mining combines machine learning, statistics, and database system together to achieve accurate results efficiently. The progressive learning style of machine learning can significantly enhance the result accuracy, especially in the AG deployment stage without a significant amount of EV-featured historical data \cite{datamag, datamag2}.
Cluster and relational analysis is used first to classify the traffic pattern, then identify influential factors to analyze historical data in South Korea. Another effective method is to use support vector machines (SVM) for more time-sensitive traffic distribution \cite{SVM}.

\subsubsection{Synthetic Model}

As the most well-known and commonly used modelling method, the synthetic model is a mathematical model that reflects the realistic physical movements of vehicles. Many classic modelling schemes fall in this category, such as the Monte Carlo simulation (MCS), queueing model, and so on.

\textbf{MCS Model:} The MCS method generates random samples based on the input data to capture the deterministic pattern in the data. \cite{mostafamodel} uses MCS to generate virtual travel distance of EV to help the operator further develop the EV electricity consumption model. The simulation principle of MCS requires a large amount of generated data to ensure the modelling accuracy, which could be time-consuming.

\textbf{Queueing Model:} The first-come-first-serve system characterizes vehicles as customers and their sojourn time as service time in the queue to model the EV on-road travelling. \cite{BCMPM} uses a Baskett, Chandy, Muntz and Palacio (BCMP) model with $M/G(n)/\infty$ queues to represent different vehicle-dense situation in road intersection by varying the $G(n)$. Queueing may be too complicated for complex traffic scenarios. Thus, this modelling method is more suitable for simple road condition such as freeway.

\textbf{Fluid Dynamic Traffic Model:} Another freeway suitable modelling method is fluid dynamic traffic model. Using the highway Poisson-arrival-location model, this method can find the arrival rate at given node on the freeway \cite{spatial}. This modelling method considers both temporal and spatial variation impacts on the traffic flow, but the application scenario is rather limited.

\textbf{Traffic Assignment Model:} As a classical mobility analysis method in the TN, this method estimates the traffic flow on the TN by assigning traffics between origin-destination (O-D) pairs in terms of travel time/alternative paths. A straightforward assignment way is to assign by probability \cite{MCS2016}. Wardrop's principles are essential in traffic assignment by concluding the user equilibrium (UE) \cite{coupledplanning} and social optimum (SO) scenarios \cite{somodel}. While SO traffic assignment provides an optimal flow distribution for the TN, UE assignment models the EV behaviour more realistically. However, UE considers that EV drivers know the complete information of the road, which may not be realistic in real life. Hence, FISK's stochastic traffic assignment model is a feasible option \cite{stochastic2018}.

\subsubsection{Simulator-Based Model}

Commercially developed simulator provides a more precise and fine-grained traffic modelling tool such as VISSIM \cite{miaomag, VISSIM}, SUMO \cite{SUMO}, PARAMICS \cite{PARAMICS}, CORSIM \cite{CORSIM} and so on. These simulators provide detailed and accurate vehicle travelling trajectory. However, they do require commercial licenses to operate while also have the potential to fail modelling the EVs due to their difference with the conventional vehicles.

\subsection{In-AG Operation Analysis}

\begin{figure}[!t]
\centering
\includegraphics[width=0.45\textwidth]{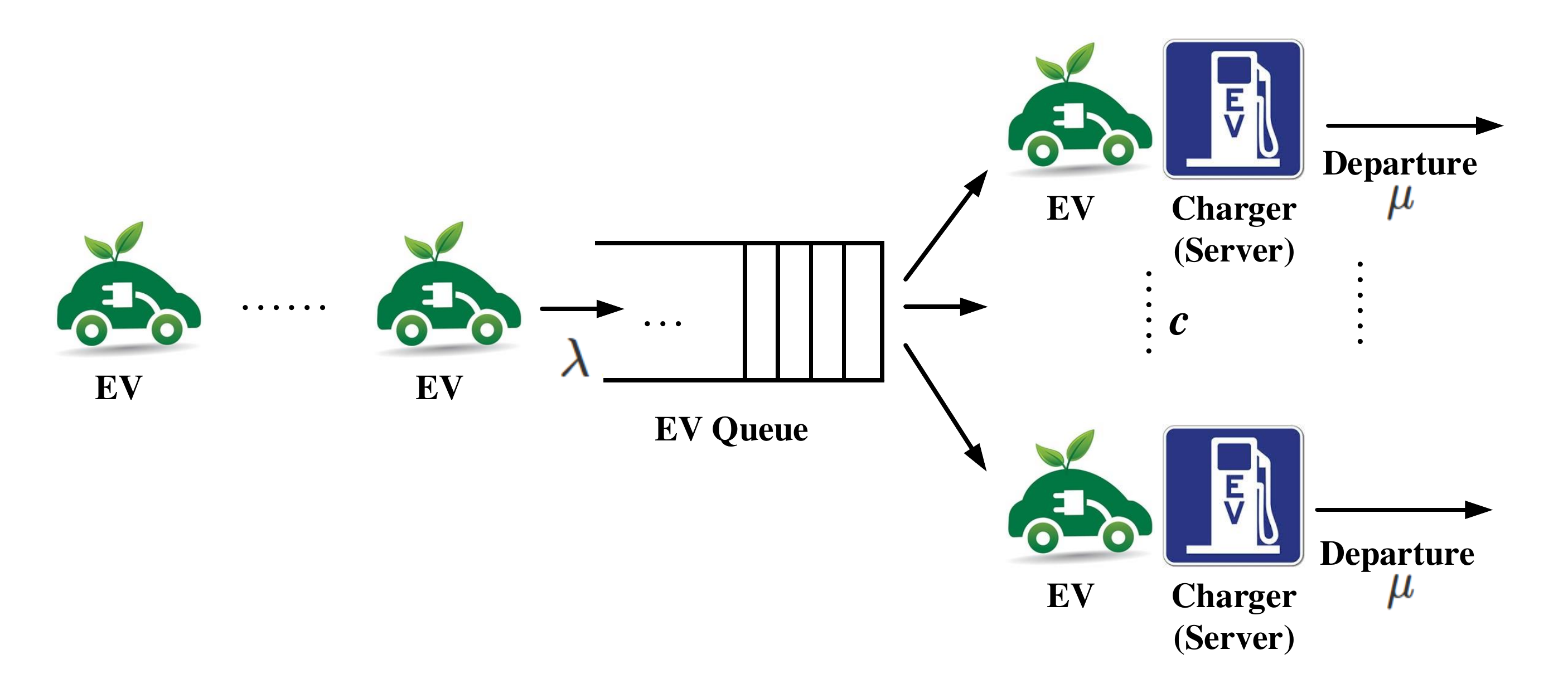}
\vspace{-0.3cm}
\caption{Illustration of EV charging queue.}
\vspace{-0.3cm}
\label{Fig_queue}
\end{figure}

With the input of EV travel pattern results, the AG service parameters (\emph{e.g.}, service capacity and waiting time) are estimated through the operation analysis. The operation is generally considered as a first-come-first-serve queueing model, which is generalized as in Fig. \ref{Fig_queue}, where EVs arrive as customers to be served. Depending on the AG properties, a variety of queue models (\emph{e.g.}, different buffer size, system size, arrival distribution, etc.) are used and reviewed below.

\subsubsection{$M/M/c$ Queue}

$M/M/c$ queue considers that EVs arrive at the AG following a Poisson process and be charged following an exponential distribution. The queue has $c$ servers with unlimited waiting space. By analyzing the operation as $M/M/c$ queue, the steady states of the queue can be obtained with respect to variant arrival and departure rates. Then, the AG service capacity can be estimated as a summation of each state probabilities multiplied by the number of EVs served in the state. Another valuable AG operation parameter is the EV waiting time, which can also be calculated using the steady-state analysis. The time-sensitivity of charging service makes the EV waiting time a non-negligible QoS metric for operation analysis \cite{MCS2016, somodel}.

\subsubsection{$M/M/c/N$ Queue}
Unlimited waiting space does not always apply to the AG, such as in the metropolitan area. In this case,  public AG has limited infrastructure space, and the waiting space can be full rather quickly. Considering this circumstance, \cite{charger} proposes an $M/M/c/N/\infty$ queue which describes the charging queue as a limited waiting space with $N$ positions in the system. With the process analysis results, this work can design the sizing scheme under the constraints of waiting time, the number of lost customers, and queue size.

\subsubsection{$M/G/c/c$ Queue}
Due to the heterogeneity of EV SoC status, the EV charging process is not strictly exponential distribution without proper charging management \cite{FRcapacity}. Hence, a more realistic model such as $M/G/c/c$ queue can be applied for AG to evaluate the EV blocking probability \cite{mgqueue}.

\subsubsection{Two Dimensional Markov Chain}
When energy storage devices are integrated to the AG to alleviate the SG loading pressure, a two-dimensional Markov Chain with finite state space is used to characterize the energy storage and EV charging operation \cite{2dmarkov, nanwcsp}. Through the analysis of the Markov Chain, the system dynamics are captured to facilitate the deployment of a network of charging stations in the city. 

\subsubsection{BCMP Queue Network}

Although multiple queue models are introduced above, the interactions between nearby AGs are also essential. To model the interaction between AGs, \cite{bcmp} utilizes a BCMP queue network where each queue in the system has a limited number of chargers and sufficient waiting space. By varying the charging prices of AGs, the traffic flow at each station varies accordingly. The interaction results can be further utilized when deploying multiple AGs.

\subsection{AG Implementation}

Based on the estimated AG service capacity, the deployment enters the implementation stage, where the siting and sizing are determined with respect to the AG's service objectives. When considered as an electric battery on the move, the EV couples the operation of both SG and TN. Thus, the allocated AG position (siting) not only needs to meet the SG power flow and economic constraints, but also the TN constraints (\emph{e.g.}, range anxiety, traffic flow maximization, etc.). Meanwhile, the sizing of the AG not only needs to satisfy the peak charging demand, but also minimize the EV waiting time. The variety of deployment objectives makes the implementation a complicated process. Next, we discuss the implementation works in terms of their objectives, as summarized in Table \ref{implementation}.

\begin{table}[!t]
\centering

\setlength{\abovecaptionskip}{0pt}
\scriptsize

\caption{OBJECTIVES FOR AG IMPLEMENTATION}
\label{implementation}

\renewcommand\arraystretch{1.5}
\begin{center}
\begin{tabular}{c|c}
	\hline
	System Perspective & Detailed Objective \\ 
	\hline
	
	\multirow{2}{*}{\centering{SG Perspective}} & Min power loss and voltage deviation \cite{123busplan, queuepv} \\
	\cline{2-2} & Min Investment and O$\&$M costs \cite{levelplan, urbantraffic} \\
    \hline

    \multirow{3}{*}{\centering{TN Perspective}} & Service radius optimization \cite{beijing, caliplan} \\
    \cline{2-2} & Accessibility improvement \cite{highwayplace, civilplace} \\
     \cline{2-2} & Max captured traffic flow \cite{korea} \\
	\hline
	
		\multirow{2}{*}{\centering{SG+TN}} & Min power loss + max captured traffic flow \cite{multiplace} \\
	\cline{2-2} & Min planning costs + max captured traffic flow \cite{multiob, stochastic2018} \\
    \hline

\end{tabular}
\end{center}
\end{table}

\subsubsection{Objective of SG}
Considering EVs as electric appliances integrated to the SG, the large-power appliances require increments in line and substation loading, which leads to increasing grid loss. For AGs in rural areas that are far away from substations, the transmission loss is also non-negligible. Thus, minimization of power loss is the main deployment objective in SG \cite{123busplan}. Besides, the active/reactive power fluctuation caused by EV integration also deviates the voltage, which needs to be minimized to guarantee the SG stability and power quality \cite{queuepv}. Meanwhile, the basic power requirements such as current line limit, transformer capacity, power balance, etc., are all included in the deployment as the problem constraints \cite{123busplan, queuepv}.

The high-power rated AG demands a significant amount of investment cost at the initial implementation stage, along with long-term operation and maintenance (O$\&$M) cost. SG operator aims to minimize the total cost during the AG operation length: investment, O$\&$M, and energy loss costs. The investment cost usually includes the installation fee of chargers and other devices, land rentals, transformer upgrading, etc. \cite{123busplan, levelplan, urbantraffic}. The O$\&$M costs include labor, electricity consumed by chargers and other devices \cite{levelplan}. When minimizing the economic cost of planning, the basic power requirements are still essential and should be included as the problem constraints.

\subsubsection{Objective of TN}

When looking at a network of AGs in the TN, the overall coverage of AGs should cover most of EV trajectories in the TN. However, as the service radius of each AG increases, the overlapping of service range between neighbouring AGs also increases, which could lead to resource redundancy \cite{beijing}. To balance between a complete coverage and minimal service redundancy, the service radius of each AG needs to be optimized. In \cite{caliplan}, the set covering problem is considered to choose the minimum set of candidate AGs that have intersections with neighbouring AGs to cover all EV routes.

A major concern of EV drivers is the fear of incomplete travel due to the limited battery capacity (\emph{i.e.}, range anxiety). Hence, the deployed AG should not only be easily accessed by distance \cite{civilplace}, but also have fast serving speed \cite{highwayplace}. This, however, demands a great budget to deploy dense, large service-capacity AGs. Thus, optimization problems that minimize EV missed trips due to battery limitation/EV waiting time are formulated under the budget constraints to improve the EV accessibility to AGs.

The service rate at the AG largely depends on its traffic flow. As a profitable infrastructure, the AG is expected to capture the maximum traffic flow at the deployed location. In \cite{korea}, the captured flow maximization is formulated as a flow refuelling location problem (FRLM) that can easily achieve the optimal result with mixed integer linear programming (MILP).

\subsubsection{Objectives of Coupled Planning}

While a maximum flow-captured location leads to profit maximization, the situation may not be feasible from the SG perspective, considering the charging impact on the SG. To balance between the charging impact and EV traffic volume, \cite{multiplace} formulates a multi-objective problem that can be transformed into a weighted single-objective problem to balance the trade-off.

In addition to the charging impact concern, the deployment cost increment is another concern for maximum flow-captured location. The high-power charging of EVs requires more investment costs (\emph{e.g.}, chargers). Moreover, O$\&$ M costs also increase accordingly, as the traffic-flow dense locations usually have higher land rents and labour demands. \cite{multiob, stochastic2018} formulates a multi-objective problem to address this issue, which can be solved by multi-objective evolutionary algorithm with decomposition (MOEA/D).

\subsection{Communication Infrastructure}

\begin{figure*}[t]
\centering
\includegraphics[width=0.8\textwidth]{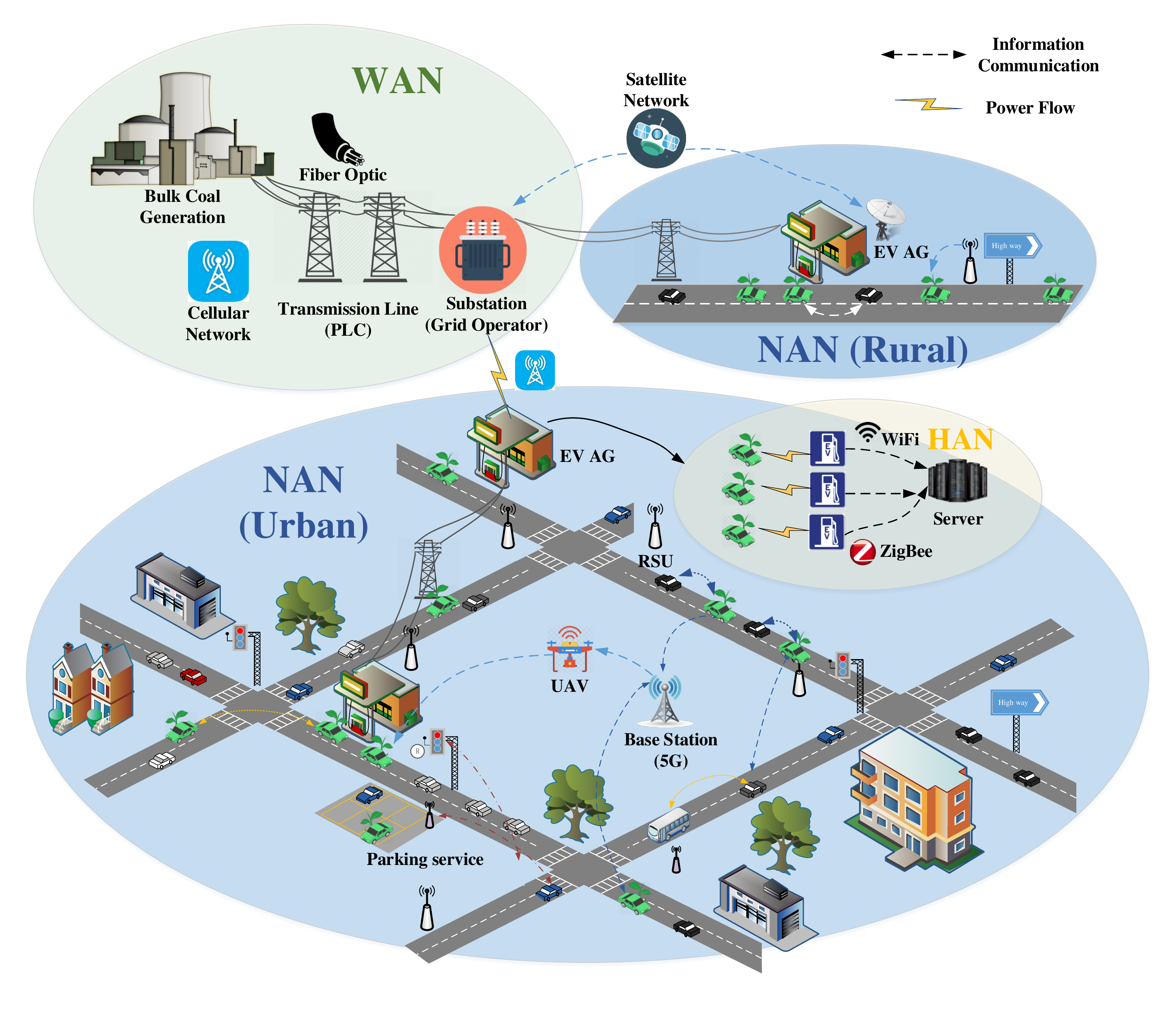}
\vspace{-0.3cm}
\caption{The heterogeneous communication environment.}
\vspace{-0.3cm}
\label{Fig_comm}
\end{figure*}

As mobile electric components in the SG, EVs can travel to different parts of the SG for charging (public/private) or V2G services. As such, the EVN communication infrastructure needs to be prepared for a variety of EVN service within a wide coverage while enabling the security and privacy of all entities in the operation. In this section, we aim to clarify the performance requirements of communication system and the development of a heterogeneous communication environment based on the existing literature, so that the deployment of communication infrastructure can be effective for EVN management.

To decide the communication technologies and networks that are adopted in the AG, we first need to clarify the communication category in the SG:

\begin{itemize}

\item Wide area network (WAN): WAN provides communication coverage from the electric utility to substation with long-distance, high-bandwidth backbone communication networks \cite{2013comm}. In the EVN, WAN provides communication between grid operators and AGs to provide timely operation statuses and commands;

\item Neighbourhood area network (NAN): NAN guarantees the communication in the power distribution system to connect between the substations and consumers. Inter-AG and AG-EVs communication fell over this category;

\item Home area network (HAN): HAN provides communication among electric appliances and smart meters at home with low-bandwidth networks. Home charging/Intra-AG communication uses HAN to enable electricity status monitor.

\end{itemize}

To enable a seamless and timely communication from grid operator to EVs, a variety of communication technologies are deployed at AG, satisfying different service requirements. Next, we clarify the performance index for data communication. Then, the heterogeneous communication environment is introduced.

\subsubsection{Performance Requirements}

In terms of different EVN service, performance requirements for data communication are introduced as follows:

\begin{itemize}

\item \emph{Latency:} Described as the delay of transmitted data, for time-sensitive services such as emergent EV charging or power ancillary services, the latency is non-tolerated. The latency requirements for these services are usually within 2 seconds \cite{2010us}, while other applications can tolerate longer time up to 5 minutes \cite{2013comm};

\item \emph{Reliability:} As a metric to describe reliable data transmission, reliability is crucial for services that heavily depend on data communication such as outage control, data monitor \cite{2013comm}. Other services can tolerate more data transfer outage;

\item \emph{Data rate:} For EV-related services, mostly EV SoC status, road, and AG conditions are the transmission contents that require a small amount of data. Thus, the data rate requirement is relatively loose \cite{2010us};

\item \emph{Throughput:} Depending on the property of transmitted data for applications, the throughput requirements vary. For example, EV charging requires a throughput of 14-100kbps \cite{2010us};

\item \emph{Carrier Frequency range:} Mobile EVs usually require high-quality and cost-effective communication among the area. For example, lower carrier frequency (below 2 GHz) helps mitigate the line-of-sight issues (\emph{e.g.}, rain fading, wall penetration, etc.) of radio signals\cite{2013comm};

\item \emph{Security:} As a cyber-physical system, the malicious attack on the transmitted data could result in severe operation interruptions or even breakdown to the SG \cite{2013comm, sgcomm1}. Providing end-to-end secure data transfer is essential to all services to not only protect the system operation but also preserve the privacy of users \cite{sgcomm1}.

\end{itemize}

\subsubsection{Heterogeneous Communication Environment}

To satisfy the EV service requirements in terms of the above indexes, a heterogeneous communication environment is demanded in the SG, as shown in Fig. \ref{Fig_comm}, where the AG geographic location and nearby traffic condition are considered. It can be seen that even for the communication between the same entities, location and traffic affect the adopted communication technologies/networks. For example, the AG in the urban area can connect with grid operator with cellular networks while the AG in the rural area with poor cellular coverage would have better performance with satellite networks. 
Next, we introduce applicable communication technologies/networks corresponding to their performance indexes, service scenarios, and communicated information, as summarized in Table \ref{comm1}.

\begin{table*}[!htbp]
\centering

\setlength{\abovecaptionskip}{0pt}
\scriptsize

\caption{APPLICABLE COMMUNICATION TECHNOLOGIES/NETWORKS}
\label{comm1}

\renewcommand\arraystretch{1.5}
\begin{center}
\begin{tabular}{|c|c|c|c|c|c|c|}
	\hline
	Category & Frequency Range & Reach & Data Rate & Latency & Application Scenario\\ 
	\hline
	Satellite Network \cite{sagsurvey, sagning, SatelliteinSG} & 1-40 GHz & Extreme large & up to 1000 Gbps & 20-280 ms & AG (rural)-SG/Inter-AG\\
	\hline
	Aerial Network \cite{confa, uavsurvey} & 2-66 GHz & Large (regional size) & 72 Mbps & Low & AG-EV (urban)\\
	\hline
    Broadband PLC \cite{IEST2010, plc2} & 1.8-100 MHz & up to 150 km & up to 200 Mbps & Low & Intra-AG/Home \\
    \hline
    Narrowband PLC \cite{IEST2010, plc2} & 3-500 kHz & up to 150 km & up to 500 kbps & Low & Inter-AG \\
    \hline	 
    Ultra-narrowband PLC \cite{IEST2010, plc2} & $\leq$ 3KHz & up to 150 km & 100 bps & Low & AG (urban)-SG\\
	\hline
	 DSRC \cite{luning, vanetgrid} & 5.850-5.925 GHz & 10-100 m & 3-27 Mbps & Medium & Inter-EV/ AG-EV \\
	 \hline
	5G (Cellular) \cite{5gsurvey1, 5gsurvey3, 5gcell} & 24-100 GHz & ~500 m/cell & up to 50 Gbps & 1 ms & AG-SG/Inter-AG/AG-EVs  \\
	\hline
     WiFi \cite{2015survey, wifirange} & 2.4 GHz, 5 GHz & 46 m (indoor)-92 m (outdoor)  & 54 Gbps & 3.2-17 ms & Intra-AG/Home\\
     \hline
     ZigBee \cite{han1, sep2} & 865MHz, 915MHz, 2.4GHz & 10-100m & 20-250 kbps & Low & In-AG/home communication\\
     \hline

\end{tabular}
\end{center}
\end{table*}

\textbf{Satellite Network:} The recent advancement of the satellite network makes it a promising communication option in the SG, expecting to achieve revenue of $\$$368 million \cite{SatelliteinSG1}. The satellite network is composed of satellites, ground stations, and network operations control centres. In terms of altitude, satellites are categorized into three types: geostationary orbit (GSO), medium earth orbit (MEO), and low earth orbit (LEO) satellites \cite{sagning, sagsurvey}. From LEO to GSO satellites, their altitudes increase accordingly from the altitudes range of 160-2000 km to 35,768 km. The long distance between satellites and ground stations incurs non-negligible data latency, especially for GSO and MEO \cite{sagning}. Recently, the advancement of LEO satellite technology presents it as a highly flexible system with relatively low latency ($\leq$ 40 ms) to transmit data in rural areas. Considering the wide coverage and low-latency of satellite network, they are very suitable for data communication in rural areas. For example, satellited network can support AG-SG and Inter-AG information transmission to keep the AG operation condition monitored \cite{sagning}. Moreover, the satellite network can support mobile data transmission for EV location/energy tracking in the rural areas where regular communication is hard/costly to reach \cite{snfv1}.

\textbf{Aerial Network}: The aerial network mainly has three infrastructures: UAVs, airships, and balloons at both high and low altitude platforms \cite{confa}. As the complementary communication to the ground network, the aerial network can be easily deployed with low-cost \cite{sagsurvey}. Their flexibility provides a variety of communication opportunities. For example, at the AG deployment stage when permanent base stations are under construction, UAVs can undertake the base station responsibilities temporarily to smooth the terrestrial network transition. They are extremely useful for data delivery between mobile EVs and AGs in the data-congested areas (\emph{e.g.}, urban) or severe communication conditions (\emph{e.g.}, shadow or interferences). During these circumstances, UAVs can deliver the EV charging/V2G requests to AGs while help advertise the AG service to mobile EVs.

\textbf{Ground Network}: The ground network consists of a variety of terrestrial communication technologies, which are introduced as follows:

\begin{itemize}

\item \textit{PLC:} Power line communication (PLC) transmits both electric power and data on the same electrical wiring, presenting as a cost-efficient data communication option for the EVN. In terms of the frequency range, PLC has different classes that are suitable for different data communication scenarios, as summarized in Table \ref{comm1}. The broadband PLC can support short-range data transmission that is very suitable for intra-AG/home communication scenario, where the AG operation status can be constantly monitored. The narrowband PLC has a longer transmission range that can support inter-AG data transmission, where the AGs can transmit their operation/pricing statuses to collaboratively work. The ultra-narrowband PLC with a 150km range and a data rate of 100 bps is suitable for communication between AGs and the SG operator to ensure regular operation in-station. Although the PLC provides a reliable and cost-efficient communication option, the noise and frequency interference of electric power transmission could degrade the communication performance \cite{surveyelectric}.

\item \textit{DSRC:} The key technology of vehicular ad hoc networks (VANETs), DSRC, has a frequency range of 5.850-5.925 GHz with a bandwidth of 10 MHz. It can achieve the data rate of 3-27 Mbps to help vehicles in proximity connect and share information \cite{luning}. By V2V and V2I communication, the technology can be used for on-road traffic/energy information dissemination \cite{vanetgrid}. In a dense vehicular environment, DSRC can help vehicles disseminate AG information (\emph{e.g.,} availability, pricing, etc.) to nearby EVs as a way to extend the AG communication coverage \cite{vFogSmartCity}.

\item \textit{Cellular Networks:} The 5G cellular network has a candidate frequency range of 24-100 GHz, with the data rate up to 20 Gbps and the mobility support up to 500 km/h according to the IMT2020 standard \cite{5gsurvey1, 5gsurvey3}. Moreover, the newly developed cellular vehicle-to-everything (C-V2X) performs superior over DSRC in terms of mobility, coverage, delay, and reliability \cite{CV2X}. The mobility-enabled, fast transmission speed of cellular networks are considered as great communication options for AG-EVs and inter-EVs scenarios, especially for time-sensitive applications (\emph{e.g.,} prior charging and regulation service). Using 5G network, AGs and EVs can connect seamlessly to monitor energy status, nearby traffic and AG conditions, and thus providing comprehensive information on-road.

\item \textit{WiFi:} WiFi is a short-distance, low-latency (3.2-17 ms) communication technology operating on 2.4 and 5 GHz frequency with the data rate up to 54 Mbps \cite{2015survey}. As the technology has been a common options for HAN, it can be easily applied to the similar scenarios of intra-AG data communication for EV charging status monitor. WiFi can also be deployed along road as access points to facilitate vehicular access to Internet on the drive \cite{XuWiFi}.

\item \textit{ZigBee:} The low-cost, power-efficient communication protocol is very suitable for HAN network and can be adopted for intra-AG data communication. The ZigBee alliance, smart energy profile 2, offers IP functionality that can be widely deployed among meters, sensors, and appliances. Thus, the interoperability between ZigBee and other IPv6 based nodes with WiFi, ethernet technologies can be effectively improved \cite{sep2}. As such, adopting ZigBee in the EVN not only provides an energy/cost-efficient option for data communication of operation monitor intra-AG, but also promotes inter-operation among different technologies.

\end{itemize}

\begin{table*}[!htbp]
\centering

\setlength{\abovecaptionskip}{0pt}
\footnotesize
\caption{HIERARCHICAL COMPUTING INFRASTRUCTURE}
\label{computing}

\renewcommand\arraystretch{1.5}
\begin{center}
\begin{tabular}{|p{50pt}<{\centering}|p{130pt}<{\centering}|p{130pt}<{\centering}|p{130pt}<{\centering}|}
\hline
	 & Cloud Computing (SG-level) & Edge Computing (AG-level) & Opportunistic Edge (EVs and UAVs) \\ 
	 \hline
	 Deployment & Deployed by IT company (\emph{e.g.,} Google, Amazon) & Deployed by local power utility company & Self-installed computer on-board \\
	 \hline
	 Server hardware & Large-scale data centers with high capacity servers & Small-scale data centers with moderate resources & Small-scale data centers with limited resources \\
	 \hline
	 Distance to end users & Long distance (regional-level) & Medium distance ($\sim$km level) & Small distance ($\sim$100 m) \\
	 \hline
	Latency & $\sim$ 5 min & $\sim$ 2 s & $\sim$ 100 ms \\
	\hline
	Application & Delay-tolerant and computation-intensive tasks that demand systematic information (\emph{e.g.,} Day-ahead planning, EV mobility learning \cite{multiaggregator, mostafarealtime,v2gcapacity2017}) & Delay-sensitive tasks (\emph{e.g.,} real-time EV charging/discharging scheduling \cite{HCFog, EVTVT, IoTFog}) and tasks offloaded from EVs (\emph{e.g.,} charging navigation) & Energy-constraint tasks, information caching and broadcasting, data collection such as environment perception \cite{vFogSmartCity, fogmag, edge1} \\ 
	\hline
     
\end{tabular}
\end{center}
\end{table*}

\subsection{Computation Infrastructure}

The EVN operation consists of multiple stages in terms of time-scale and security concerns. For example, SG-level operation is conducted in the day-ahead power market while EV charging/V2G service operation in-AG is real time. Moreover, considering the privacy-sensitive of EVs and AGs, detailed operation information (\emph{e.g.,} charging rate, service admission, etc.) should be processed locally. Therefore, a hierarchical computing infrastructure is demanded to fulfill EVN computing tasks at different levels. In this section, we first clarify the computation requirements for the EVN operation. Then, we introduce the computing infrastructure for the EVN and their detailed properties, as summarized in Table \ref{computing}.

\subsubsection{Performance Requirements}

As a hierarchical system, EVN demands a variety of computation service at different level. Therefore, in terms of data size, context, location, and energy demand, performance requirements for the computation service vary \cite{MECsurvey, MECComSurvey}, and the main metrics are introduced as follows: 

\begin{itemize}

\item \textbf{Latency} - Described as the delay of data computation, for the service that demands fast response such as charging navigation for mobile and energy-constraint EVs, the requirements of communication and computation latency are usually within 2 seconds \cite{2010us}. In such a case, allocating the service to a edge server that is closer to the user end with sufficient computation capacity could effectively satisfy the latency requirement.

\item \textbf{Storage} -  data storage capacities of cloud and edge server are quite different as cloud nodes have overwhelmingly larger capacity than edge nodes. Therefore, for the services that heavily depend on data analysis (\emph{e.g.,} distributed energy prediction and EV mobility analysis), the tasks should be performed on the cloud while the edge nodes store and analyze real-time data that will be updated regularly.

\item \textbf{Energy efficiency} - Another key metric, especially when the user devices are used for computation is energy efficiency. For computation intensive tasks, they can be offloaded to nearby edge servers or cloud servers to reduce the user's energy consumption.

\item \textbf{Security} - Malicious attacks on hierarchical computation could lead to severe operation interruption and privacy leakage. For computation tasks that are offloaded to cloud nodes, unauthorized access to the node could lead to privacy leakage, confidential issues and even disrupting operation. Meanwhile, edge nodes are vulnerable to external security threats regarding server access and channel breach \cite{MECsurvey}. Therefore, well-designed countermeasures for potential threats are in urgent demand.

\end{itemize}

\subsubsection{Cloud Computing at the SG}

The deployment of cloud computing at the SG level provides a sharing environment for operators to store/share information, build/develop applications, and operate system with online software \cite{cloudSG}. Supported by industrial products, the grid operator can perform different computing tasks efficiently \cite{CloudPowerGrid}. Google Cloud SQL provides a platform to perform energy scheduling and communication service as software as a service(SaaS) \cite{GoogleCloud}; Amazon elastic compute cloud can provide data storage by virtual machine for the grid operator as infrastructure as a service (IaaS) \cite{EC2}. The grid operator can also utilize servers such as CloudFusion to build and develop programming models as platform as a service (PaaS) \cite{CloudFusion}.

For cloud computing at the SG level, the operator only collects descriptive data (\emph{e.g.,} AG charging/discharging demand, power statuses at the SG buses) to monitor the SG operation. Meanwhile, detailed information such as on-road EV charging demand and vehicular traffic is stored and processed in edge nodes. As the cloud node has a global overview of the SG, it can undertake delay-tolerant computing tasks that require systematic information \cite{multiaggregator, mostafarealtime,v2gcapacity2017}. For example, the cloud node can allocate power to each AG in the day-ahead power market while the AG performs the computing task of charging scheduling within the allocated power range using its local computing resource. Moreover, when edge nodes accumulate a large number of computing tasks, they can also offload the computing-intensive and delay-tolerant tasks to the cloud node.

\subsubsection{Edge Computing at AGs}

For the AGs that are equipped with computation, storage, and communication devices, they are great candidates to perform as edge nodes near end users \cite{HCFog, EVTVT, IoTFog}. AGs perform as crucial bridges in the hierarchical computing architecture that interact with different entities.

In the EVN, AG is not only a computing node, but also an energy interface between EVs and the SG. Within the AG, the AG operator computes the detailed operation process (\emph{e.g.}, charging rate, service admission, etc.). This computing task is time-sensitive and privacy sensitive, and it cannot be distributed to other edge nodes. Similarly, upon service agreement with the SG, the V2G scheduling tasks should be computed within the AG.

The AG can undertake computing tasks from its nearby EVs that have limited computing capacities. Computing tasks such as charging location selection and travelling navigation can be partially offloaded to the edge nodes. In this case, EVs only execute data collection (\emph{e.g.}, sensing battery status) and final selection part (\emph{e.g.}, choosing its preferred AG) locally, while computation-intensive components (\emph{e.g.}, energy charging optimization/comparison of different AGs) are performed at edge nodes \cite{MECComSurvey}.

For AGs deployed in dense vehicular areas, computing tasks can accumulate very fast and computing tasks can be offloaded to other edge nodes with computing capacities or the remote cloud node. This approach can effectively enhance the computing resource utilization in the EVN. However, the server selection and cooperation needs to be well-designed considering the EV mobility, spatial and temporal computing task arrivals \cite{MECComSurvey}.

\subsubsection{Vehicles and UAVs as Opportunistic Edge Nodes}

The mobility and sensors on-board empower vehicles and UAVs with location-based sensing and processing capabilities so that they can work as opportunistic edge nodes. However, the edge computing and sensing responsibilities also lead to challenges such as energy-inefficiency and data heterogeneity. In the following, we discuss their computing roles and challenges in detail.

\begin{itemize}

\item \textbf{UAVs} - Owing to their fast deployment and mobility, UAVs can either collect data from sensors deployed at rural areas or sense data in the territory that is hard to reach \cite{UAVMCS}. Moreover, UAVs can flexibly offload communication/computing tasks from end users as opportunistic edge nodes \cite{uavsurvey}. Since UAVs are energy-constraint, the balance between sensing/computing efficiency and energy consumption is a critical issue \cite{MECComSurvey}. To perform a sensing/computation-intensive tasks before its deadline could incur high energy consumption beyond the UAV range. In this case, the tasks should be offloaded to nearby UAVs and AGs.

\item \textbf{Vehicles} - Equipped with both sensors (\emph{e.g.,} camera, GPS, radar, etc.) and processing units, mobile vehicles can cluster together to provide location-based services, which is considered as mobile crowdsensing. In the introduced computing architecture, these clustered vehicles work as opportunistic edge nodes to sense, process, and upload the real-time data, contributing to an efficient EVN operation \cite{EdgeComputingMCS, Peng}. For instance, the real-time vehicular travelling data not only helps the operator track the on-road traffic condition \cite{smartcity}, but also contributes to accurate EV travelling pattern prediction \cite{EVMCSIEL}. Their on-board computers empower them as edge computing nodes to perform a variety of computing tasks such as energy data pre-processing and battery management. However, the stability and resource allocation of edge nodes face great challenges, and their utilization as IaaS still needs further development \cite{VehicularFogInfras}.

\end{itemize}

\subsection{Security of Information Infrastructure}

As the foundation of the EVN, the SG encounters numerous security and privacy challenges owing to its increasing IoT device deployment, heterogeneous communication environment, and high-reliability requirement \cite{Ning2}. For readers who are interested in the SG security survey, please refer to \cite{securitydata, security2, shsecurity}. Meanwhile, the vehicular network also has its unique security and privacy concerns in terms of location privacy, availability, authenticity, and so on. There have been extensive works in this area, and readers can refer to \cite{VANETsecurity, vanetse} for details. Further, the adoption of edge computing and cloud computing brings new challenges to the system operation, regarding data access control, privacy preservation, integrity, reliability of the collected data \cite{securecloudedge}. There have been extensive research works conducted on security issues of cloud computing \cite{cloudSG, cloudsecurity1, cloudsecurity2} and edge computing \cite{SecureEdgeReview}.

\subsubsection{Security Requirements for EVN}

As the intersection between the smart grid, vehicular network, and hierarchical computing, the EVN has its unique security and privacy challenges. For instance, the payment between EVs, AGs, and the SG requires high security and mutual authentication for energy/information transaction \cite{securitydata}. Moreover, the communicated information, including EV trajectory, battery status, and grid electricity demand, is highly sensitive and requires secure communication protocol to prevent eavesdropping and malicious attacks during the communication process \cite{IoEV}. To ensure that effective security mechanisms are designed for the EVN operation, we elaborate the key security requirements below.

\begin{itemize}

\item \textbf{Availability} - The EVN operation demands strong robustness, and therefore, the availability of communication channel/computation space (\emph{e.g.}, CPU) is the key security metric to enable robust operation under the presence of faulty and malicious condition.

\item \textbf{Confidentiality} - The high sensitivity of communicated data (\emph{e.g.,} AG and grid operation condition, EV battery status and preference, etc.) requires high confidentiality to prevent unauthorized access and disclosure.

\item \textbf{Integrity} - To prevent both physical and cyber security attacks (\emph{e.g.,} energy theft), the communicated data should be protected against unauthorized modification or destruction.

\item \textbf{Authentication}- Authentication between EVs, AGs, and grid operator is essential in the EVN operation to provide a trust relation among communication entities. In terms of different contexts (\emph{e.g., charging or discharging}), the authentication scheme needs specific design.

\item \textbf{Access control} - A hierarchical system such as the EVN requires a well-designed access control management to ensure that each entity can be assigned with appropriate data access authority. For example, AGs can access local information such as EV status to ensure smooth operation while the grid operator has a higher access authority for system-level operation.

\item \textbf{Privacy preservation} - The privacy preservation of EVs is essential as many communicated data are closely related to EV owners' personal information, and privacy leakage could lead to severe safety concerns. Sensitive data such as battery status, location information, and behaviour preferences could be used carefully to interpret the EV usage pattern upon disclosure.

\end{itemize}

\subsubsection{Security Threats of EVN Operation}

The SG operator along with its equipped cloud computing node is vulnerable to be attacked by external attackers while its controlled nodes (such as AGs and EVs) are honest-but-curious, and they may snoop on the personal information of data owners, causing privacy leakage. A external/internal attacker may launch the following attacks to disrupt the EVN operation:

\begin{itemize}

\item \textbf{Denial of Service attack} - By jamming the communication channel with data packets or consuming CPU memory resources with specific request packets, authorized entities such as EVs could be prevented from accessing services (\emph{e.g.,} AG information request) \cite{VANETsecurity, SecureEdgeReview}.

\item \textbf{Jamming attack} - The attacker can disrupt the communication channels with strong signals to prevent information communication between different entities \cite{IoEV}. Jamming attack occurs in various operating scenarios, \emph{e.g.}, intra-vehicles to interrupt the data exchange and condition monitor between battery sensors and vehicular computers/drivers \cite{IoEV}.

\item \textbf{Malware attack} - When a malicious software is installed into OBUs, RSUs, and even more high-authority units such as SCADA, attackers can easily penetrate the EVN to disrupt normal operation \cite{IoEV}.

\item \textbf{Broadcast tampering attack} - If insider attackers broadcast fake information (\emph{e.g.}, power shortage or AGs are fully occupied), the correct information from trusted operators could be overridden, which leads to degraded EVN performance.

\item \textbf{Black/Gray hole attack} - Black and gray hole attacker behaves similarly to regular entities, and (selectively) drop relaying packets during transmission \cite{VANETsecurity}. For example, when AG advertisements are transmitted via V2V communication, a corrupted vehicle could drop the advertisement and result in the delay of time-sensitive information.

\item \textbf{Eavesdropping} - Unauthorized nodes detect/extract confidential information from the protected data. For example, attackers could extract a household activity pattern through collecting their electricity consumption profile.


\item \textbf{Sybil attack} - Sybil attackers cover themselves under multiple identities (\emph{e.g.,} charging-demand EVs) in different positions to inject fake information (\emph{e.g.,} AG reservation and incorrect battery status) to the EVN. In such a case, the AG operator will be misled by the received information and lead to disrupting operation such as AG congestion or EV energy shortage.

\item \textbf{GPS spoofing} - The attacker can generate fake GPS signals that have stronger signals than the trusted ones to EVs, deceiving their location information. The threat becomes severe when EVs are in rural areas and have limited battery energy for travelling.

\item \textbf{Unauthorized access} - The attacker can illegally access the SDN controller, cloud computing controller or even the SG controller and then manipulate the system operation, which could lead to severe system failure \cite{cloudsecurity1}.

\item \textbf{Modification attack} - The attacker can alter the messages sent among different entities (\emph{e.g.,} Inter-EVs/AGs-EVs/AGs-SG operator) to interrupt system operation \cite{V2Vauthentication}. For example, EVs that sent charging requests to nearby AGs may receive messages stating that nearby AGs are fully occupied whereas they are available for charging.

\item \textbf{Masquerading attack} - Using the stolen passwords to enter system control platform, the attack can broadcast false messages and disrupt the regular operation.

\item \textbf{Replay attack} - The attacker records the information exchanged among EVs, AGs, and SG operator, and continuously re-inject the record data back to the network \cite{V2Gmag}. This behaviour will confuse the operator, especially in emergency conditions such as micro-grid outage or AG congestion.

\end{itemize}

As the EVN operation demands sensitive data of users, the collection, transmission, processing, and sharing of data could be disclosed to unauthorized entities and expose the privacy of data owners. The users' privacy can be divided into four aspects: identity privacy, data privacy, location privacy, and usage privacy \cite{JBSecurity}.

\begin{itemize}

\item \textbf{Identity privacy} - The identity of a user includes name, address, telephone number, and license number, \emph{i.e.}, any information that can link to a specific user. When the information is sent to AGs/SG operator for authentication, the user identity is vulnerable to be disclosed.

\item \textbf{Data privacy} - The data privacy could be leaked to an untrusted party if the party has unauthorized access to the data storage or eavesdrops during data transmission. By analyzing the data, sensitive information could be disclosed.

\item \textbf{Location privacy} - For EVs moving along the road for charging, discharging, and navigation services, their location privacy preservation is critical. With the collected location information, an attack is able to identify the user's location, travelling trajectory, etc.

\item \textbf{Usage privacy} - The disclosure of usage privacy refers to the usage pattern (\emph{e.g.,} EV charging profile, EV preference, community electricity profile, etc.), which severely violates the user's privacy.

\end{itemize}

To satisfy the above security requirements and tackle the security threats, extensive research works have been conducted on EV charging scheduling and V2G technology, which will be discussed in Sections IV-B (3) and V-C (3).

\section{EV CHARGING SCHEDULING}

Different from AG deployment which considers the EV service demand fluctuation from a long-term perspective, the charging scheduling considers real-time factors such as real-time EV charging demand and impact. Moreover, in terms of the control signals and methods of different charging scheduling, the EVN information management requires different technologies and networks to guarantee efficient and timely operation.

In this section, we first identify challenges of EV charging scheduling. Then, existing works of EV charging scheduling are reviewed in terms of energy and information management.

\subsection{Charging Scheduling Challenges}

The EV charging scheduling demands coordination among mobile EVs, AGs, and the SG from all three aspects of energy, communication, and computation. It faces the following challenges:

\begin{itemize}

\item The AG operators aim to maximize the financial profit and operation performance, whereas AGs are under the energy constraints of infrastructure and SG planning. The balance between the operation objective and energy constraint is challenging;

\item From EVs' perspective, they aim to maximize the service QoS with minimal costs. When AG operators schedule EV charging, it is challenging to achieve global or near optimal results considering the selfish behaviours of EVs;

\item For EV charging to exchange data among the SG, AGs, and mobile EVs, each entity has different data update frequency and control range. It is crucial and challenging to identify applicable communication technology to achieve effective scheduling;

\item Considering the variety of data content, server hardware, computation capacity in the EVN, the allocation of EV charging scheduling tasks is critical and challenging. 

\end{itemize}

\begin{figure*}[!t]
\centering
\includegraphics[width=1\textwidth]{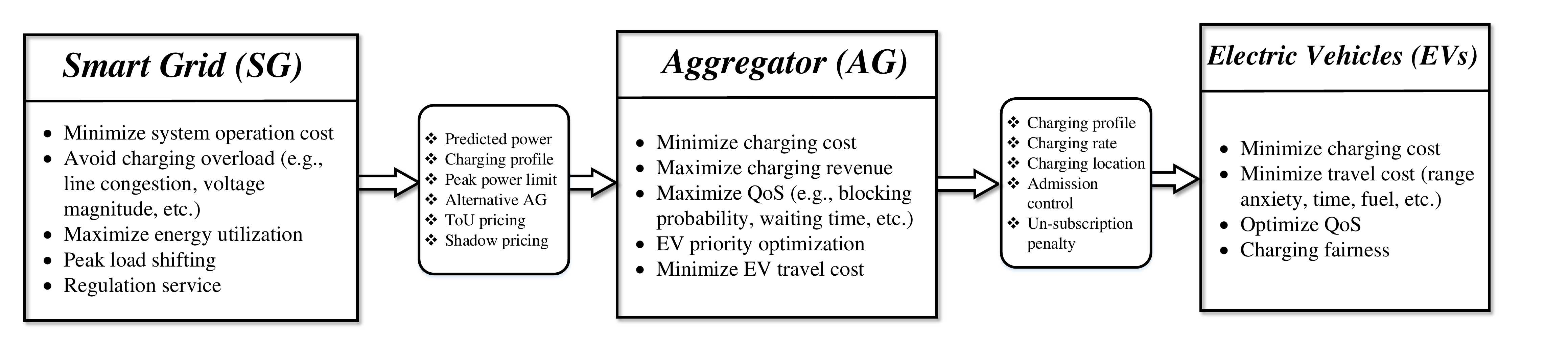}
\vspace{-0.3cm}
\caption{Entity interaction with respect to objectives.}
\vspace{-0.3cm}
\label{Fig_interaction}
\end{figure*}

\subsection{Energy Management for EV Charging}

Based on the existing research works, the summarized scheduling objective and control methods between SG, AGs, and EVs is shown in Fig. \ref{Fig_interaction}. It can be seen that the SG mainly tries to mitigate the EV charging impact, especially during peak hours. When EVs are flexible for scheduling, they can also be used to facilitate the ancillary service. The SG scheduling objective is transmitted to AGs as either direct power allocation command/preference or in-direct control signal such as electricity price.

On the other end of scheduling, EV drivers aim to accomplish the charging tasks with optimal economic cost and high QoS. While some EVs are comfortable following the AG commands, others behave selfishly and need incentive to follow the scheduling.

Considering scheduling objectives of both SG and EVs, AGs need to enable its economic income while satisfying both power constraints and QoS. In terms of the objectives and entity interaction, we categorize the EV charging scheduling works into central control and indirect control, as summarized in Table \ref{chargingcoordination}.

\begin{table*}[!htbp]
\centering

\setlength{\abovecaptionskip}{0pt}
\scriptsize

\caption{Control methodology of EV charging scheduling}
\label{chargingcoordination}

\renewcommand\arraystretch{1.5}
\begin{center}
\begin{tabular}{|c|c|c|c|}
\hline
	Control Category & Control Variable/Method & Scheme Objective & Solution/Technique\\ 
	\hline
	 \multirow{9}{*}{\centering{Central control}}& \multirow{5}{*}{\centering{Charging profile}} & Min operation cost + Max energy utilization \cite{mostafarealtime} & General algebra modeling system\\ 
      \cline{3-4} & & Min EV charging cost \cite{adaptive} & DC MILP	\\
       \cline{3-4} & & Min day-ahead energy cost +charging time \cite{2016twostage} & Convex optimization \\
        \cline{3-4} & & Min day-ahead operation cost + emission \cite{EV1} & PSO \\
         \cline{3-4} & & Max charging income \cite{EV4} & Relaxed convex optimization \\
	 \cline{2-4}& \multirow{2}{*}{\centering{Charging rate}} & EV charging priority optimization \cite{realtimemarket} & Linear programming \\
	  \cline{3-4} & & Max revenue + QoS (rejection rate) \cite{contract2018} & Two-step iterative algorithm \\
	 	\cline{2-4} & Penalty for EV un-subscription & Max revenue \cite{multiaggregator} & Linear programming \\
	 	\cline{2-4} & Charging location & Min travel cost + Max charging power \cite{vanet1} & Time-coupled MILP \\
	 	\hline
	 	\multirow{9}{*}{\centering{Indirect control}}& Power-varying pricing & Load shifting \cite{valleyfill} & Convex optimization \\
	 	\cline{2-4} & Shadow pricing & SG: min power cost; AG: min charging cost \cite{congestion} & Convex optimization \\
	 	\cline{2-4} & Congestion pricing & Neutralize wind farm generation \cite{windint} & VMP calculator design \\
	 	\cline{2-4} & \multirow{2}{*}{\centering{Charging price}}& Load shifting \cite{ganlingwen} & Asynchronous optimal decentralized charging \\
	 	\cline{3-4} & & Min overall electricity cost + power deviation cost \cite{Info2} & Linear programming \\
	 	\cline{2-4} & Two-stage non-cooperative game & AG: min power charing cost + EV: max EV charging energy \cite{game2018} & Iterative best-response algorithm \\
	 	\cline{2-4} & Hierarchical game & AG: max revenue + EV: min economic and travel cost \cite{2018navigation} & Particle swarm optimization \\
	 	 \cline{2-4} & \multirow{2}{*}{\centering{Stackelberg game}} & AG: max revenue and min QoS + EV: min QoS and travel cost \cite{stackelberg2d} & Best response algorithm \\
	 	 \cline{3-4} & & AG: max revenue + EV: min travel and waiting cost \cite{stackelbergmulti} & Subgame perfect equilibrium \\
\hline

\end{tabular}
\end{center}
\end{table*}

\subsubsection{Central Control}

When EVs are centrally controlled by AGs, they usually provide AGs with their required charging amount and charging duration via communication. Upon receiving the EV information, an optimization problem is formulated at the AG to achieve the optimal operation by scheduling the EV charging profile/rate/admission, etc.
There are mainly two objectives when EVs are centrally controlled: achieving financial optimality or achieving optimal charging performance.

\textbf{Financial Optimality:} As a profitable public infrastructure, AG aims to minimize its operating cost or maximize its charging income to achieve financial optimality \cite{multiaggregator, mostafarealtime, adaptive, contract2018}.

\begin{itemize}

\item \textit{Operating cost minimization} - The operation cost consists of two parts: electricity cost and power loss cost.  The main portion of the operation cost, \emph{i.e.}, electricity cost, comes from drawing electricity from the SG, which is arranged day-ahead based on the AG predicted charging demand \cite{2016twostage, EV1, EV3}. In the real-time market, the SG also uses electricity prices to control the AG power demand indirectly \cite{multiaggregator, mostafarealtime}. The power loss of EV charging is another part of operation cost, and is proportional to the overall charging energy \cite{mostafarealtime}.

\item \textit{Charging income maximization} - AG charging income includes the charging profit and penalty fees of EVs that fail to arrive at scheduled AG. Intuitively, with more EVs arriving at AGs, charging profit increases accordingly \cite{EV4}. However, the incoming EV charging demand needs to be leveraged with constraints of the SG and EVs \cite{mostafarealtime, adaptive, contract2018, EV4}.

\end{itemize}

To optimize the AG financial performance, the following constraints are considered during the operation:

\begin{itemize}

\item \textit{Active/reactive power}: Active and reactive power equality should be satisfied at every power bus to ensure energy balance in the SG;

\item \textit{Voltage/current}: Magnitudes of both voltage and current should be within their regulated ranges, which are determined by the SG operator. For example, typically, voltage magnitude should be between $95\%$ and $105\%$ of its nominal magnitude;

\item \textit{Battery status}: For each EV, the charged energy should be within its feasible range (\emph{e.g., less than its battery capacity}). Meanwhile, the overall charging demands in the AG should be within its charging capacity.

\end{itemize}

Considering the non-linear properties of the problem constraints (\emph{e.g.,} power equality, current constraints, etc.), the scheduling problem needs to be relaxed as an MILP or convex problem to find an optimal solution \cite{vanet1, adaptive, EV2}.

\textbf{Optimal Charging Performance:} To evaluate the charging performance, there are many metrics. First, energy utilization percentage helps the SG and AGs evaluate the occupancy of the deployed AGs to see the infrastructure pay-off long-term \cite{mostafarealtime, vanet1}. Second, from the EV drivers' perspective, the in-AG waiting time and congestion condition directly affect EV charging experience \cite{2016twostage, vanet1}. Without proper scheduling, a congested AG could lead to a high customer dissatisfaction, which results in reputation degradation. As an approach to improve the QoS, AG can consider the priority of EV charging to satisfy the urgent charging needs \cite{realtimemarket}.

Although some works sorely consider the AG performance optimum as the scheduling objective  \cite{realtimemarket}. In the real-life scenario, AG needs to balance between the financial cost and performance while guaranteeing the SG constraints.

\subsubsection{Indirect Control}

Considering most EVs tend to behave selfishly by charging the vehicles at their convenience, central control could fail. In a more realistic scenario, AG achieves its scheduling objective by controlling signals such as charging price. EVs make their individual decision based on the posed price and their objectives, which is defined as indirect control.

Many research works apply indirect control method by either designing pricing schemes \cite{congestion, ganlingwen, windint, valleyfill} or using mathematical frameworks (e.g., game theory) \cite{2018navigation, stackelberg2d, stackelbergmulti, game2018} to model the hierarchical structure of EV charging. Next, we discuss these two methods in detail.

\textbf{Designed Pricing Scheme:} Presently, the SG already implements time-of-use (ToU) price to regulate electricity consumption during different periods. However, this flat pricing scheme does not consider real-time system status and could lead to unexpected line congestion/overload \cite{stackelberg2d}. Thus, dynamic pricing schemes are designed to adjust price corresponding to real-time charging condition. The pricing scheme usually consists of two parts: designing the pricing function and modelling the user response to the price.

One common pricing function is power-varying pricing where consumers are charged for energy and power usage \cite{valleyfill}. This pricing scheme can be very effective when the controller tries to shift EV charging load temporally. Another similar pricing method called congestion pricing provides users who are willing to pay more with larger resource \cite{windint}. This price-incentive method can also facilitate EVs to provide ancillary service for integrated RESs. To enable a globally optimized control results, \cite{ganlingwen} designs a decentralized algorithm to guarantee the optimal result convergence, even with asynchronous information exchange. \cite{Info2} considers a day-ahead pricing scheme for residential demand response to schedule EVs and renewable sources, aiming to minimize both residential electricity bill and fluctuating load profile. \cite{congestion} designs a shadow pricing scheme that considers the Lagrange multiplier of the optimization problem as the shadow price to iteratively adjust the price converging.
The modelling of user response to price is studied either by formulating an optimization to satisfy the EV's objective \cite{ganlingwen} or projected user behaviour \cite{valleyfill, windint}.

\textbf{Game Theory:} Game theory is a very suitable mathematical model that can simulate the interaction between EVs and AGs. For example, EV charging can be considered as a competitive game since each EV would like to obtain its maximum profit selfishly \cite{game2018}. If an AG has the lowest charging cost, EVs will choose it. Meanwhile, some EVs also prefer AGs with good QoS (e.g., waiting time, blocking probability) \cite{stackelberg2d, stackelbergmulti}, and easy accessibility \cite{stackelbergmulti}. For EVs with similar routes or the same ownership, they can cooperate to bargain with AGs, forming an evolutionary game \cite{2018navigation}.

Different responsibilities of EVs and AGs result in a hierarchical game structure. With one AG scheduling EVs, the game can be formulated as a Stackelberg game. The AG plays as the leader that first acts (\emph{e.g.}, optimizes its cost), and poses its charging price to EVs. Then, EVs, as followers, make their decisions (\emph{e.g.}, choose whether or not to enter the AG) accordingly \cite{stackelberg2d, stackelbergmulti}. On the other hand, AGs belonging to different companies can form a non-cooperative game to maximize their charging revenues. Meanwhile, multiple prices are posed to EVs for their decision making \cite{game2018}. The hierarchical game is solved by the best response algorithm to achieve equilibrium at each level. However, the uniqueness and optimality of the equilibrium depend on the proposed utility function property.

\begin{table*}[!htbp]
\centering

\setlength{\abovecaptionskip}{0pt}
\footnotesize

\caption{DATA COMMUNICATION FOR EV CHARGING}
\label{comm2}

\renewcommand\arraystretch{1.5}
\begin{center}
\begin{tabular}{|p{40pt}<{\centering}|p{40pt}<{\centering}|p{100pt}<{\centering}|p{120pt}<{\centering}|p{140pt}<{\centering}|}
	\hline
	Category& Scenario & Comm. Requirement & Comm. Content & Applicable Technologies/Networks \\ 
	\hline
    \multirow{2}{*}{SG-AG} & Day-ahead & High-security & charging profile, predicted power capacity & PLC, satellite \cite{plc2, SatelliteinSG} \\
     \cline{2-5} & Real-time & Low-latency, high-security & Locational marginal price, peak power shifting & Cellular, satellite \cite{cellular2, jsacoffload} \\
     \hline
	\multirow{2}{*}{\shortstack{AG-EV\\(Central)}} & Intra-AG & Low-latency, low data rate & Charging profile, charging rate & PLC, ZigBee, WiFi \cite{plc2, zigbeeEV, 2015survey}\\
	\cline{2-5} & Mobile EVs & Mobile, low-latency, reliable & Charging location,
admission, penalty & Cellular, DSRC, aerial \cite{urllcVN, miaomag, aerialoffload} \\
	\hline
	AG-EV (in-direct) & Mobile EVs & Mobile, extreme low-latency & Charging price & Cellular, aerial \cite{IoV, uavIoV}\\
	\hline
     
\end{tabular}
\end{center}
\end{table*}

\subsection{Information Management for EV Charging Scheduling}

Information is indispensable in achieving effective EV charging scheduling, as data communication, computing, and security are crucial steps in terms of securely exchanging system information and optimizing the management. In this section, we first introduce the data communication for EV charging in terms of operation scenario, applicable technology, and communication content. Then, the related works on computing application for EV charging are reviewed and categorized in terms of computing methodology. Finally, the security-related efforts on EV charging are introduced in terms of securing EV payment, distributed security scheme, and location privacy preservation.

\subsubsection{Data Communication for EV Charging}

Data communication for EV charging is hierarchical and demands exquisite technology selection for suitable scenarios. Based on the existing works, we summarize the data communication requirement, content, and application, as shown in Table \ref{comm2}.

\begin{itemize}

\item \textbf{Communication for the SG Signal:}
In terms of the energy markets, the SG has day-ahead and real-time markets \cite{ieso}. The day-ahead market provides a dependable view of the next day available supply and anticipated demand by estimating historical data and other factors. The SG can transmit their preferred charging profile, predicted power capacity to AGs day-ahead so that AGs can perform their local scheduling tasks accordingly. The day-ahead data communication does not have high requirements on latency and data rate, but does require a highly secure link to protect the SG from cyber and physical attacks. Wired communication technologies such as PLC can be used for the day-ahead communication \cite{plc2}. For the AGs in the rural area where the wired infrastructure can be costly, the satellite network can be applied \cite{SatelliteinSG}. 

\setlength{\parindent}{2em} The real-time market frequently updates the locational marginal price to match the supply and demand, usually every five minute \cite{ieso}. The peak power shifting tasks are also transmitted to the AGs to help alleviate the loading pressure. Compared with the day-ahead market, data communication of the real-time market needs a low-latency, high-security technology to guarantee that the AGs can receive and perform the scheduling on-time. Cellular networks are great candidates for this type of communication \cite{cellular2}. For areas with poor cellular coverage, the satellite network is also a potential option \cite{jsacoffload}.

\item \textbf{Communication for Central Control:}
When AGs receive the information from the SG, they perform the charging scheduling scheme in central control mode. The control signal of the scheme usually consists of two types: the intra-AG plugged-in signals (\emph{e.g.}, charging profile, charging rate) and the AG-EV signals (charging location, admission, penalty). For the first type of signal, as EVs are usually stationary and connected to the AG, wired communication such as PLC is a cost-efficient and convenient option \cite{plc2}. To timely adjust the charging profile of EVs, low-latency data transfer is required. Meanwhile, as the data is transmitted as small-size packets, the data-rate requirement is usually low. Hence, wireless technologies that are adopted in HAN such as ZigBee and WiFi are also applicable \cite{zigbeeEV, 2015survey}.

\setlength{\parindent}{2em} Another type of signals that are usually transmitted between AGs and mobile EVs include information of AG condition and EV charging. Based on the real-time market information, AGs could frequently change their control signals to achieve their scheduling objective. Thus, mobile, low-latency, and reliable technologies are needed. For EVs in the urban areas, cellular networks are great option \cite{urllcVN} while DSRC and aerial networks can also be applied in the congested and rural areas \cite{miaomag, aerialoffload}.

\item \textbf{Communication for In-direct Control:} The interaction between AGs and mobile EVs are through in-direct control signals such as pricing. Considering the process of AG sending signals and EVs responding to the signal, the communication latency doubles compared to the centrally controlled case \cite{vanetcharging}. Therefore, URLLC is demanded to enable the data stay valid during the communication process (\emph{e.g.}, and EVs do not miss the destined AG on-the-move) \cite{IoV}. 5G network, along with aerial networks can be adopted to accomplish this task \cite{uavIoV}.

\end{itemize}

\subsubsection{EV Charging Computing}

The energy management of EV charging demands a platform for data storage and analysis, that is computing infrastructure. In this section, we review related works of cloud/edge computing on EV charging scheduling, as summarized in Table \ref{computing2}.

\begin{table*}[!htbp]
\centering

\setlength{\abovecaptionskip}{0pt}
\footnotesize

\caption{Computing Application on EV Charging Scheduling}
\label{computing2}

\renewcommand\arraystretch{1.5}
\begin{center}
\begin{tabular}{|c|p{100pt}<{\centering}p{80pt}<{\centering}|p{100pt}<{\centering}|c|}
	\hline
	 Research work & Cloud & Edge & Objective & Technique and solution\\
	\hline
	D. A. Chekired \emph{et al.} \cite{cloudEV} & SG: data share + computing & \ding{55} & Calendar scheduling & Priority assignment scheduling \\
	\hline
     Y. Chen \emph{et al.} \cite{DRCload} & SG: incentive computing & RES-DG: load operation & Min overall operation cost & Binary linear programming \\
     \hline
      Y. Cao \emph{et al.} \cite{MECbigdata, EVMEC} & \ding{55} & Vehicle/UAVs: information caching & Charging recommendation & Process flow design \\
      \hline
      S. S. Shah \emph{et al.} \cite{vFogSmartCity} & \ding{55} & Vehicles: edge computing & Enhance computation + reduce transmission delay & vFog framework \\
      \hline

\end{tabular}
\end{center}
\end{table*}

\begin{itemize}

\item \textbf{Cloud computing for EV charging} - The cloud computing platform provides IaaS and SaaS for EV charging scheduling at the grid operator level. A data storage device (or warehouse) can be deployed at the operator side to support different EV charging services by architecture design \cite{cloudSG53} or management optimization \cite{cloudSG54}. On the other hand, utilizing the accessed data and online software can significantly enhance the scheduling performance of EV charging \cite{cloudEV, DRCload}. In \cite{cloudEV}, EVs update their information and create charging/discharging calendars on the cloud, sharing with the grid operator. Based on the information, the grid operator schedules EV charging calendars with the priority assignment scheduling. The data sharing and computing can be more efficient when the computing entity is within a community formed by distributed energy resources such as EVs, and storages \cite{DRCload}. In this case, the grid operator provides incentives to each community to smooth the load fluctuation while minimizing their operation costs. Cloud computing has been a mature technology that can be efficiently applied in cyber-physical system such as the SG. For readers who are interested in cloud-aided charging, please refer to \cite{cloudSG}.

\item \textbf{Edge computing for EV charging} - To alleviate the communication burden on the backbone network, AG information such as AG waiting time and charging recommendation can be cached at the edge nodes such as vehicles and UAVs. Then, moving EVs opportunistically access the cached data at the nodes, and make their charging reservations through edge nodes according to the designed process flow in \cite{MECbigdata, EVMEC}. Moreover, edge computing also enhances the computation and communication performance on-road by shortening the distance between the computing devices and EVs \cite{vFogSmartCity}.

\end{itemize}

\subsubsection{Security of EV Charging}

As suggested in \cite{securitydata}, one of the fundamental security issues of EV charging is ensuring a secure payment scheme. The EV payment system has three unique features: two-way transaction for charging/discharging, privacy-preserving, and traceability of transaction with consent \cite{payment}. To preserve the EV privacy while enabling the transaction traceability, \cite{payment} proposes a novel payment system considering attacks of location privacy infringement, fraudulent statement, slandering, and hiding. \cite{Info1} proposes that EVs buy anonymous coins from the bank for charging payment at AGs to ensure anonymous payment and authentication. With the proposed hierarchical authentication scheme that uses hashing/Exclusive-OR cryptosystems, attacks against driver privacy and payment can be effectively prevented. 

As more and more EVs travel on-road, distributed security schemes can be more efficient than centralized schemes. As an emerging paradigm, blockchain provides an efficient and secure interaction model among entities and has been widely adopted in data communication and computing.  With the blockchain, the charging scheduling can be conducted without a centralized authority, which is suitable for V2V charging scenario. In \cite{BCC35}, a novel peer-to-peer energy trading model based on consortium blockchain is proposed for secure transaction and privacy preservation of EV charging. A contract-based energy blockchain is proposed in \cite{blockchainCharging} so that EVs can publicly share transaction records without relying on a trusted intermediary. Moreover, the roles of EVs in the EVN as both energy customers and computing devices can be fully exploited using blockchain. \cite{securecloudedge} introduces energy and data coins as the proof of energy contribution amount and the proof of data contribution frequency, respectively. Using energy and data coins as authentication operators, the work proposes secure computing schemes to achieve data transparency and traceability.  In \cite{Ningblock}, blockchain is applied to record energy transactions among distributed EVs, enabling a secure incentive scheme for EVs. 
The authentication of EV charging is also essential in both EV-AG and EV-EV charging scenarios. To prevent the man-in-the-middle attack during V2V charging, \cite{V2Vauthentication} uses a mutual challenge/response protocol with Diffie-Hellman key exchange.

Compared to gasoline-based vehicles, EVs stop at AGs for charging more frequently due to their limited battery capacities and therefore, their travelling patterns are easier to capture \cite{payment, Info3}. While privacy preservation for charging in-AG has been discussed in \cite{payment, Info1}, the linkability between EVs and AGs should also be reduced to preserve EV location privacy \cite{Info3}. In \cite{Info3}, power routers are introduced to obfuscate the linkages between charging events. The proposed obfuscation scheme effectively thwarts location inference and therefore enhance the location privacy.

\section{VEHICLE-TO-GRID (V2G) Technology}

In addition to converting electrical energy for mechanical propulsion, the large-capacity battery of the EV can feed energy back to the SG to improve the stability and reliability of the grid \cite{surveyelectric, ecosurvey}. This reversing power transfer from EVs to SG is called vehicle-to-grid (V2G), which is supported by both rapid EV commercialization and battery development. As predicted by \cite{global}, by 2030, there will be 56 million EVs in the vehicle market, while the battery capacity increased to 120kWh on average. As 90$\%$ of vehicle operation time is spent stationary in the parking lots \cite{concept1}, they can be effectively used by the SG when parked. Furthermore, the lithium-air technology has the potential to upgrade the battery energy density to 2000-3500Wh/kg, compared with the current density of 100-150Wh/kg \cite{battery}, making the battery lighter and energy-efficient.

To understand the V2G potential applications in the SG, we first analyze the role of V2G in the electricity market, along with its management challenging. Corresponding to the challenges, existing works on V2G energy management are reviewed with respect to the service objective, while information management for V2G is presented in terms of data communication, service computation, and security.

\subsection{Roles and Challenges of V2G Services in the Electricity Market}

According to \cite{ecosurvey}, the electricity market has four sub-markets in terms of control methods: base-load power, peak power, spinning reserve, and regulation markets. The mobility and large fleet size of EVs make them flexible and fast-response energy storages, which is suitable for the latter three sub-markets. However, the mobility and range anxiety issues of EVs could impede the smooth role transition of EVs from loads to energy suppliers, while the information management of V2G service can be complicated owing to heterogeneous service. The following aspects should be addressed when managing V2G services:

\begin{itemize}

\item The mobility of EVs brings uncertainty and potential delay to the V2G service. While in some cases, the affect can be mitigated by online generations, the dissatisfaction of time-sensitive service (\emph{e.g.,} regulation) could lead to system-level reliability issues;

\item The discharging of EVs not only incurs range anxiety concerns for the EV users, but also raises their concerns about battery degradation. Therefore, proper V2G scheduling needs to consider EV travelling demands and depth-of-discharge (DoD) status;

\item As V2G services have heterogeneous energy demand and response requirements, it is challenging to configure data/command transmission among the SG, AGs, and (mobile) EVs;

\item The computing management from cloud to edge and opportunistic edges is crucial in the V2G scenario as the service is closely related to the stability and energy balance in the SG. Therefore, it is crucial to design effective computing scheme under various operating scenario.

\end{itemize}

\subsection{Energy Management for V2G Services}

\begin{figure}[!t]
\centering
\includegraphics[width=0.5\textwidth]{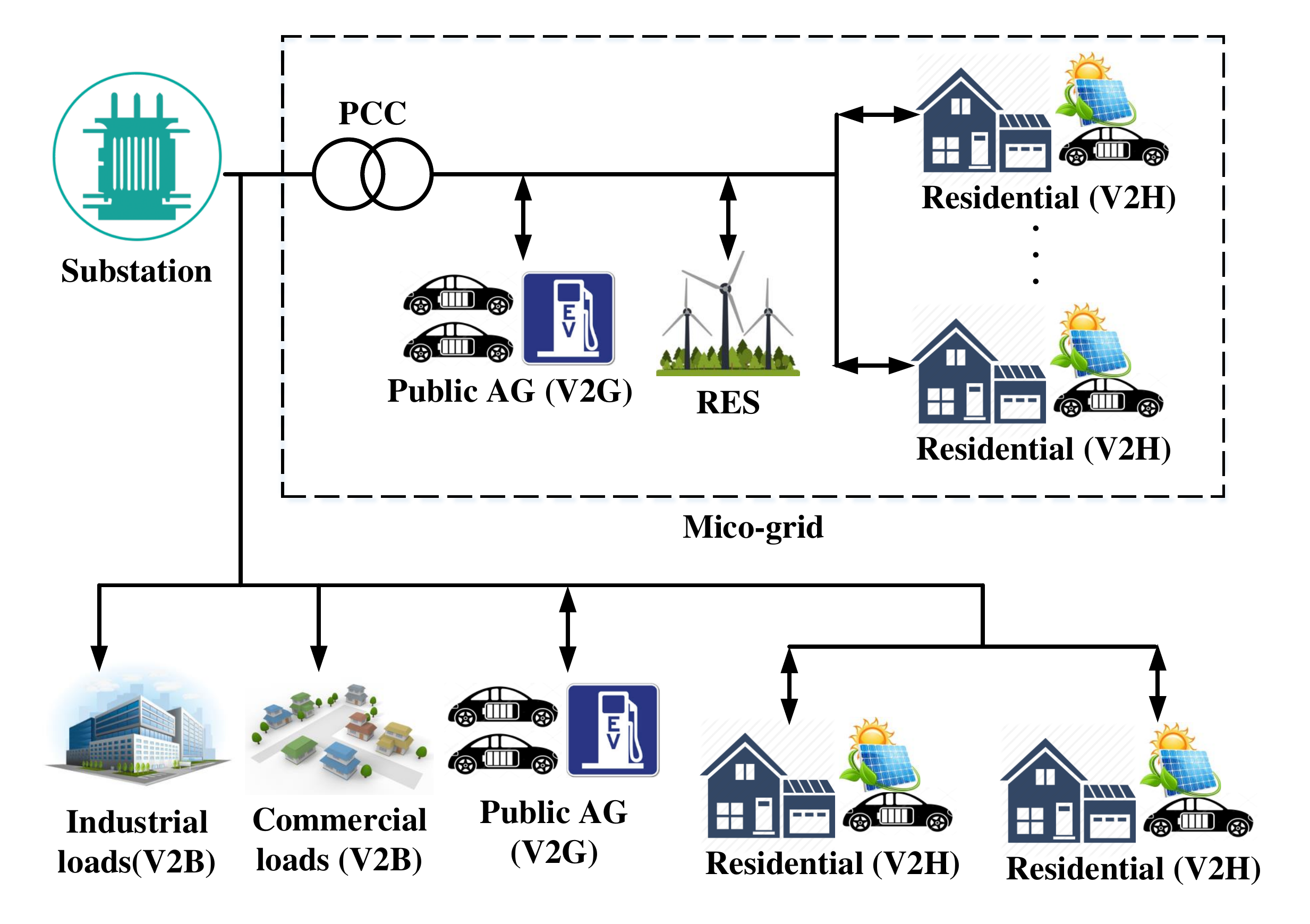}
\vspace{-0.3cm}
\caption{An overview of energy management of V2G.}
\vspace{-0.3cm}
\label{Fig_v2g}
\end{figure}

Depending on the scheduling objective, there are three types of V2G working modes, as shown in Fig. \ref{Fig_v2g}. When EVs are connected at home to provide energy for home appliances with RES, vehicle-to-home (V2H) mode is engaged. When EVs are driven to workplaces (industrial loads) or shopping malls (commercial loads) and connected to the parking lot chargers, they can perform discharging service in vehicle-to-building (V2B) mode. In the peak hour, vehicle-to-grid (V2G) is performed at AG to avoid overload. 

It can be seen that V2G has a variety of services that range from one residential home to the large-scale microgrid. Next, we discuss the V2G services in detail in terms of their objectives, as summarized in Table \ref{v2g}.

\begin{table*}[!htbp]
\centering

\setlength{\abovecaptionskip}{0pt}
\scriptsize

\caption{METHODOLOGY FOR MODELLING AND OPTIMIZATION IN V2G}
\label{v2g}

\renewcommand\arraystretch{1.5}
\begin{center}
\begin{tabular}{|c|c|c|c|}
	\hline
	Service Category & Application Scenario & Scheduling Objective & Solution/Technique \\ 
	\hline
	 \multirow{8}{*}{\centering{DSM}}&  \multirow{7}{*}{\centering{Load flatten}} & Min EV energy cost \cite{energymini} & Stochastic inventory theory \\
	 \cline{3-4} &  & Demand balance among district \cite{v2gmobilenetwork} & Complex network synchronization \\
	 \cline{3-4} &  & Min EV travel+charge cost \cite{nanTVT} & Convex optimization \\
	 \cline{3-4} &  & Min microgrid cost +peak load deviation \cite{energystorage} & Branch and brand pool search \\
	 \cline{3-4} &  & Min microgrid cost +peak load deviation \cite{islanded} & MILP \\
	 \cline{3-4} &  & Min EV discharge cost \cite{workplace} & Convex optimization \\
	 \cline{3-4} &  & Max SG+EV rewards \cite{mobilityv2g} & Value iterative algorithm \\
	 \cline{2-4} & \multirow{6}{*}{\centering{Outage management}} & Min Battery degration+electricity cost \cite{2018resident} & Augmented epsilon-constrain \\
	 \cline{3-4} &  & Min operation+interruption cost \cite{V2H} & MILP \\
	 \cline{3-4} &  & Max energy utilization \cite{reliability2016} & Quadratic programming \\
     \cline{3-4} &  & Max critical load restored + Min number of switching operation \cite{MAS} & Multi-agent operation \\
       \cline{3-4} &  & Min electricity cost - battery cost \cite{ParkinglotOutage} & MILP \\
        \cline{3-4} &  & Max critical load restored - allocation cost \cite{ElectricBus} & Stochastic nonlinear programming \\
	 \hline 
	 \multirow{5}{*}{\centering{RES Integration}}&  \multirow{3}{*}{\centering{Energy utilization}} & Min electricity cost+ waiting time \cite{RESconf} & Natural aggregation algorithm \\
	 \cline{3-4} &  & Min electricity cost \cite{homeRES} & Stochastic optimization \\
	 \cline{3-4} &  & Min EV energy loss \cite{mobileflow} & Minimum-cost flow \\
	 \cline{2-4} & \multirow{2}{*}{\centering{Reliability assessment}} & Simulation based assessment \cite{resco} & Matlab \\
	 \cline{3-4} &  & Min reliability cost + EV cost+ power loss \cite{windpower} & Probabilistic load flow \\
	 \hline
	  \multirow{6}{*}{\centering{Regulation service}}&  \multirow{3}{*}{\centering{Centralized control}} & Max service revenue-operation cost-peak load deviation \cite{v2gcapacity2017} & MINLP \\
	  \cline{3-4} &  & Min EV expense \cite{adesign2} & Quadratic programming \\
	  \cline{3-4} &  & Max EV revenue+fairness \cite{FRprofit} & Convex optimization \\
	  \cline{2-4} &  \multirow{3}{*}{\centering{In-direct control}} & AG: meet regulation demand + EV: meet SoC demand \cite{hierregulation} & Fuzzy logic control \\
	  \cline{3-4} &  & AG: max revenue+EV: min service cost \cite{distributedFR} & Iterative pricing-based control \\
	  \cline{3-4} &  & Max AG+EV revenue \cite{regulationgame} & Iterative best response algorithm \\
	  \hline

\end{tabular}
\end{center}
\end{table*}

\subsubsection{Demand Side Management (DSM)}
When EVs are connected to the AG to discharge, the variant loads at peak hour can be supplied by EV energy locally to flatten the load profile. The EV scheduling in DSM mode is usually centrally controlled by the grid operator. Thus, V2G scheduling can be formulated as an optimization problem that aims to minimize the scheduling cost. Generally, the scheduling costs consist of SG operation cost, peak load variation cost, interruption cost, and EV costs. Depending on different DSM services, the scheduling objective (\emph{i.e.,} cost composition) changes accordingly, which will be discussed in Sections V-A (1)-(3). Meanwhile, V2G scheduling is constraint by the SG power operation and EV discharging as follow:

\begin{itemize}

\item \textit{Active/reactive power}: Similar to the EV charging scheduling, V2G scheduling should always guarantee the equality of acitve/reactive power of each power bus to enable energy balance in the SG;

\item \textit{Voltage/current}: The magnitude of node voltage and current should always be within its regulated range to guarantee the power quality;

\item \textit{AG energy balance}: At every AG, the generated electricity and EV discharged energy should match the local EV charging demand and other loads to enable the effectiveness of V2G;

\item \textit{EV discharging}: The V2G power at each AG should fit its discharging power standard. For every EV, its discharged energy should be within the battery feasible range (\emph{i.e.,} between its minimal required SoC and its battery capacity);

\item \textit{EV status}: While EVs undertake roles of energy consumers and suppliers in the EVN, they cannot charge and discharge simultaneously. Therefore, the charging/discharging statuses of EVs need to be regulated efficiently;

\item \textit{Other constraints}: EV mobility requirements such as its spatial scheduling flexibility, on-road traffic condition, EV/AG QoS requirements need to be considered in the V2G service.

\end{itemize}


\textbf{Load Flatten:} A main V2G service in DSM is to flatten the peak load by supplying the surplus demand with EV energy. Consider the long-time duration of peak load, a large number of EVs will be connected for service. However, each EV has different arrival SoC status and required departuring SoC, which demands flexible scheduling and monetary incentive to stimulate them discharging \cite{energystorage, mobilityv2g}. Moreover, the battery degradation incurred by discharging needs to be consider. Thus, the battery usage cost is incorporated in the scheduling objective, and it consists of three parts:

\begin{itemize}

\item \textit{EV charging/discharging costs}: The operator pays EVs to charge or discharge to flatten the loaf profile in terms of the energy amount EVs provide \cite{islanded, workplace};

\item \textit{Battery degradation}: When EVs are discharged upon request, additional battery degradation cost is paid as the battery degradation compensation. The cost is a function of discharging power and battery depth-of-discharge (DoD) \cite{NanGlobecom};

\item \textit{Mobility cost}: When EVs are considered as mobile energy storages to transfer energy from energy-dense areas to energy-sparse areas, the cost incurred by EV mobility should be considered \cite{nanTVT}. Factors such as travel distance, delay, energy loss, and on-road traffic condition need to be considered when calculating the mobility cost \cite{mobileflow}.

\end{itemize}

%

Another part of the load flatten scheduling cost is caused by demanding surplus energy from generation systems when the local energy generation and EV discharging cannot match the load demand \cite{energymini, v2gmobilenetwork}. In addition to the above two economic scheduling objectives, the grid operator also needs to optimize the system performance. As the violent variation of power load leads to high ramp-up rate at the generation side, the power deviation at peak hour needs to be minimized \cite{mobilityv2g, energystorage}. The peak load deviation cost is calculated as the quadratic function of the power deviation between real loads and expected loads \cite{mobilityv2g, energystorage}.

%

\textbf{Outage Management:} Apart from load flatten, EVs can be deemed as emergent energy suppliers when the local grid encounters outage. When an outage happens at home, the main V2G scheduling objective is to supply the prior appliances (\emph{e.g.}, fridge, heating) \cite{V2H, reliability2016}. Therefore, the interruption of appliances should be mitigated. The interruption cost can be calculated as the summation of interruption cost of home appliances. For different appliances with various priorities, the interruption cost is different. Service for high-priority appliances will lead to a higher interruption cost. While minimizing the interruption cost, the grid operator also needs to ensure the energy balance between emergent suppliers and home demands.

EV application in outage management is essential, especially considering microgrid will be the main component of the SG. Different from V2H, EVs that are integrated in microgrid serve a large amount of energy in a flexible manner. The energy requirement can be met by either using a large number of parked EVs \cite{MAS, ParkinglotOutage} or using large-capacity EVs such as electric bus \cite{ElectricBus}. The EV-microgrid management mainly aims to maximize the critical loads restored by local energy sources and EVs \cite{MAS, ElectricBus}. Due to the randomness of fault location and EV battery status, the outage management can adopt probabilistic model/stochastic programming to characterize the EV service capacity and fault condition \cite{ElectricBus, ParkinglotOutage}. From simulation in \cite{ParkinglotOutage}, it is shown that the participation of EVs can effectively reduce the load interruption, guarantee the normal operation of an islanded microgrid, and reduce line loss.


While some operators tend to schedule DSM in an economical way to minimize the system cost, the V2G impacts on EVs also need to consider EV requirements in the problem formulation. Thus, the discussed objectives can be chosen and formulated as a single/multi-objective optimization problem. The multi-objective optimization problem can be reformulated as a weighted-sum single objective problem and obtain the Pareto optimality for different scheduling scenarios \cite{energystorage, 2018resident}.

\subsubsection{RES Integration}

As the environmental concern of greenhouse gas emission arises, an increasing number of RES-DGs will integrate to the SG for green energy supplement. The weather/geographic-dependent power generation leads to violent output fluctuation, which could jeopardize the system reliability \cite{resco, windpower, RESparking}. As potential energy storages that mitigate the RES fluctuation, EVs can be connected to the SG together with RES to store their surplus energy and compensate for the energy gap when necessary \cite{review2015}.

While the modelling of EV mobility has been explored in the previous section, stochastic RES output also needs to be studied. An intuitive method is data-based modelling, where the historical RES data are extracted to build a probabilistic model \cite{RESparking}. Other similar approach such as MCS generate random samples for capturing the output distribution \cite{RESconf}. The synthetic model that transforms probabilistic real-life environment and generation process into mathematic representation is adopted in \cite{resco}. However, the stochastic properties of RES-DGs and EVs inevitably bring prediction error and uncontrollable energy imbalance to the SG, demanding well-designed V2G scheduling.

RES can be integrated into the SG as an additional supply option in each residential home, which provides consumers with two advantages. First, using RES energy at peak-hour reduces the purchase of retail electricity, thus, reducing the overall electricity cost \cite{homeRES}. Second, RES also helps supply home appliances during outage \cite{RESconf}. Similar to the DSM case, an optimization problem for RES integrated with V2G can be formulated. The objective is to minimize the electricity cost and maximize RES energy utilization. Some additional constraints on the home appliance operations (\emph{e.g.}, priority, operation interruption, etc.) can be added correspondingly to help achieve a cost-efficient and effective home operation.

On the other hand, for RES deployed as large-power rating farm, delayed demand response could lead to energy waste. In such case, EVs can transfer the RES energy to the energy-demanding area \cite{mobileflow}. The V2G scheduling then becomes an energy-constraint traffic assignment problem that is solved to achieve a timely and energy-efficient result.

The reliability concerns incurred by RES and EV integration is also an essential research issue. Assessment metrics such as loss-of-load/loss-of-energy probabilities are used to evaluate the system performance \cite{resco}. Through simulation, it is shown that proper integration of RES and EVs improve the reliability \cite{resco}. The reliability is also considered as the V2G scheduling objective and can be characterized as the product of energy imbalance of RES and interruption cost \cite{windpower}.


\subsubsection{Regulation Service}

Regulation services refer to the automatic generation control which fine-tunes the frequency and voltage of the system through balancing between supply and demand \cite{ecosurvey}. The stochastic property and time-sensitivity make the regulation a highly suitable market for EVs \cite{FRprofit}. Recently, the capability of reactive power compensation with V2G is investigated in \cite{reactiveshiyan}, raising the interest in voltage regulation. However, most existing works focus on frequency regulation scheduling, which will be our main topic. The frequency regulation service is further divided into two types: regulation up (RU) and regulation down (RD). RU service increases the power generation from a baseline level while RD decreases it from the baseline. In terms of the service payment, it has two components: capacity price $p_\mathrm{c}$ and electricity price $p_\mathrm{e}$. Capacity price is paid for reserved power that is available for RU/RD while electricity price is paid for the power that is delivered back to the grid \cite{FRprofit}.

To provide frequency regulation service, AGs usually need contracts with the SG to settle down their service capacities. The EV mobility and limited service capacity make the service capacity estimation an essential research subject. Consider the memoryless property between EVs on-the-move, \cite{FRcapacity} models the EV service provision as a queue network with $M/M/\infty$ queues representing charging, RU, and RD respectively. Another estimation method used in \cite{v2gcapacity2017} is through an AG model to record and forecast EV frequency regulation capacity every quarter-hour. Based on the signed contract, AGs can provide frequency regulation service to the SG by either centrally controlling EVs or schedule the service with control signals. As the RU could raise concerns about battery degradation and range anxiety, the main objective of a centrally controlled scheduling focuses on maximizing the EV revenue \cite{v2gcapacity2017, adesign2, FRprofit}. The EV revenue is composed of three components:

\begin{itemize}

\item \textit{Capacity/electricity revenue}: EV capacity revenue is the product of EV reserved energy and capacity price. Electricity revenue is the multiplication of EV discharged energy and electricity price. To enable the fairness of service allocation, an SoC-related proportional allocation factor $\alpha_\mathrm{i}$ can be introduced as the coefficient of the revenue. $\alpha_\mathrm{i}$ should be determined so that a low-SoC EV provides less RU service to avoid battery depletion and a high-SoC EV discharges more;

\item \textit{Battery degradation cost}: This cost is caused by discharging EV for RU service, which is calculated as the product of the degradation price and discharged energy;

\item \textit{Penalty fee}: When EVs leave AGs and their desired SoC statuses are not reached, the grid operator would have to pay penalty fees to EVs.

\end{itemize}

The frequency regulation scheduling is often formulated as an optimization problem, subject to power flow, EV power and energy constraints, similar to the DSM case. With pre-defined power flow limit, the problem can be simplified as a convex problem that can be solved in polynomial time \cite{adesign2, FRprofit}, while other problems with AC power flow limit is mixed-integer non-linear programming problems. To fulfill frequency regulation with hundreds of EVs in minutes, central control encounters potential issues of high computation complexity and high communication overhead.

To alleviate the computation and communication overload at one central point, the AG can assign some service flexibility to the EV level so that EVs make their service options (\emph{e.g.}, V2G power, SoC status, etc.). In some cases, EVs sign contracts with the AG to confirm their service capacity. Then, upon receiving the service requests from the SG, AGs send the required capacity to EVs while EVs at the lower level adjust the power rating to fulfill the task \cite{hierregulation}. In other cases, EVs provide frequency regulation service stimulated by high service price. Hence, a hierarchical game framework is a common approach \cite{distributedFR, regulationgame}: both AGs and EVs aim to maximize their revenue while accomplishing the regulation task with price as the control signal in-between. The game framework is usually solved by iterative best response algorithm, and achieve the equilibrium optimality depending on the payoff function formulation.



\begin{table*}[!htbp]
\centering

\setlength{\abovecaptionskip}{0pt}
\footnotesize

\caption{DATA COMMUNICATION FOR V2G}
\label{comm3}

\renewcommand\arraystretch{1.5}
\begin{center}
\begin{tabular}{|p{60pt}<{\centering}|p{60pt}<{\centering}|p{100pt}<{\centering}|p{110pt}<{\centering}|p{120pt}<{\centering}|}
	\hline
	Service & Scenario & Comm. Requirement & Comm. Content & Applicable Technologies/Networks \\ 
	\hline
    \multirow{3}{*}{\shortstack{Load flatten\\(DSM)}} & SG-AG & High-security, low-latency & Load profile, incentive & Cellular, satellite \cite{networkedEV, SatelliteinSG} \\
     \cline{2-5} & AG-EV (parked) & High-security & EV condition (\emph{SoC, availability}) & PLC, ZigBee, WiFi \cite{plc2, han1, HAN2}\\
     \cline{2-5} & AG-EV (mobile) & Mobile support, low-latency, high-data rate & EV condition, navigation & Cellular, aerial, DSRC \cite{uavIoV, interworkingsurvey} \\
     \hline
	\multirow{2}{*}{\shortstack{Outage control\\(DSM)}} & Islanded microgrid &  Low-latency, low-energy, reliable & Energy providing requests & PLC, ZigBee \cite{plc2, HAN2}\\
	\cline{2-5} & Individual home & Low-energy, reliable & Home appliance scheduling & PLC, ZigBee \cite{plc2, HAN2}\\
	\hline
     \multirow{2}{*}{\centering{RES integration}} & Home integrated & High-security, reliable & RES energy condition  & PLC, ZigBee, WiFi \cite{plc2, ZigBeePLC} \\
	\cline{2-5} & Large farm & Mobile, extreme low-latency & Regional energy status & Cellular, aerial \cite{cellularrenewable}\\
	\hline	
	 \multirow{2}{*}{\centering{Regulation service}} & Contract & Low-latency, high-security & Service duration and capacity & PLC, WiFi \cite{FRprofit, softwareWiFi} \\
	\cline{2-5} & Incentive & Extreme low-latency, reliable, high-security & Incentive, service confirmation & Cellular network \cite{2013comm}\\
	\hline	
	
\end{tabular}
\end{center}
\end{table*}

\subsection{Information Management for V2G Services}

The V2G service can only be effective when data communication, computation are securely and timely performed. Compared to EV charging management, where the AG operator focuses more on analyzing the intra-AG operation, the nature of V2G service (\emph{e.g.,} outage management, regulation, etc.) injects more stochastic properties to the management. Therefore, data communication and computation needs to be carefully considered under different V2G scenarios. In this section, we first discuss the adopted communication technology and transmission content of V2G service in terms of service type. Then, existing works on computation for V2G service are reviewed. Finally, security schemes on payment, authentication, and privacy preservation are categorized and reviewed.

\subsubsection{Data Communication for V2G}

For EVs discharging to serve different electricity sub-markets, the transmitted data and communication requirements differ in terms of the electricity sub-market. Based on the existing works, we summarize the communication requirements, content, and application, as in Table \ref{comm3}.

\begin{itemize}

\item \textbf{Load Flatten (DSM):}
The peak load occurred during peak hours requires ramp-up of bulk generations, in which case, leading to both economic and power losses. To manage the peak load, the grid operator estimates the load fluctuation in real-time and sends the load amount that needs to be flatten to the AG along with any desired economic incentive. As the load frequently fluctuates in real-time, the information needs to be sent timely for AGs to respond. Hence, a low-latency, high-security communication technology is required, such as cellular and satellite networks \cite{networkedEV, SatelliteinSG}

\setlength{\parindent}{2em} For the EVs that provide load flatten service, they are either parked at home/parking lots or used as energy storage devices to transmit power on the move. For stationary EVs in the parking lots, they communicate with AGs about their demanded SoC requirements and service time to help AGs arrange the V2G service for each EVs. Considering the privacy of the information, the data transfer needs to be highly secured \cite{sgcomm2}. Similar requirements can also be applied to stationary EVs at home as for now the electric appliances at home are also participants in the load flatten service. Thus, high-security wired communication such as PLC and wireless communication such as ZigBee, WiFi are all great options \cite{plc2, han1, HAN2}.

\setlength{\parindent}{2em} For mobile EVs that are used as mobile energy storage to transfer energy from surplus energy areas to sparse areas for load flatten, the communication requirement is more strict as the time schedule can be tight regarding the service emergency \cite{nanTVT}. Not only are the energy-related data transmitted between EVs and AGs, sometimes navigation data are also required to guarantee timely operation result. Hence, a mobile-support, low-latency, high data rate communication technology is required, such as cellular,  aerial networks, and DSRC \cite{uavIoV, interworkingsurvey}.

\item \textbf{Outage Management (DSM):}
When outages happen in the distribution system, connected EVs can be used as temporary generators for emergent usage. When the outage happens at the PCC, the outage detections would respond quickly and break the PCC for local power safety, forming an islanded MG. Thus, connected EVs in the micro-grid are scheduled to support the local area, which requires not only quick scheduling but also low-latency, low energy, and reliable communication technologies such as PLC \cite{plc2} for AG-EVs and ZigBee for home \cite{zigbeemicrogrid, HAN2}.

\setlength{\parindent}{2em} When the outage happens in a small area, EVs are mainly considered as private generator at home to provide individual energy demand. The scheduling aims to provide energy to high priority appliances which are mostly plugged-in. In this case, reliable and low-energy communication such as PLC and ZigBee can undertake the data communication tasks \cite{plc2, HAN2}. 

\item \textbf{RES Integration:}
RES is integrated to the SG as either distributed generator at load side or large-size generators in the generation subsystem. For RES integrated at load side, EVs behave as the energy buffer to store or consume RES generated energy when needed. Moreover, other home appliances also participate in RES integration. To guarantee a reliable and balanced energy condition at home, high-security and reliable communication are needed (\emph{e.g.}, PLC, ZigBee, WiFi) \cite{ZigBeePLC}.

\setlength{\parindent}{2em} When RES is integrated to the SG as large-size generators (\emph{i.e.}, farms), its accessibility is less convenient. EVs are now considered as mobile energy storages that transmit energy from RESs to other energy-demanded areas to smooth the RES integration \cite{pingyi}. In this case, EVs need to stay connected with AGs to keep up with the RES and road conditions. The randomness of RES makes this integration tasks time-sensitive, requiring extreme low-latency and mobile communication technologies such as cellular and aerial networks \cite{cellularrenewable, aerialIoT}.

\item \textbf{Regulation Service:}
Regulation service such as frequency regulation is closely related to the SG stability. The frequency control system has three layers: primary, secondary, and tertiary control, whose response time increases from seconds to 15 minutes. The frequency regulation is usually in the secondary control range whose response time is at minute-level. As discussed before, part of EVs settle down contracts with AGs to regulate their frequency regulation service time and capacity. In this case, the communication content between two entities is just service duration and required capacity. The limited response time demands low-latency, high-secure communication such as PLC, WiFi \cite{FRprofit, softwareWiFi}.

\setlength{\parindent}{2em} EVs without pre-set contracts can also provide V2G service when motivated by monetary rewards from AGs. However, the process is longer as real-time agreements need to be reached between EVs and AGs. This process consists of incentive broadcasting, EV decision transmission, and AG response. The high-reliability, low-latency requirements of regulation service demands URLLC of 5G network to help accomplish the process \cite{URLLCMag}. In URLLC, satisfying both latency and reliability requirements is still a challenging issue that needs to be further investigated.

\end{itemize}

\begin{table*}[!htbp]
\centering

\setlength{\abovecaptionskip}{0pt}
\footnotesize

\caption{Computing Application on V2G Scheduling}
\label{computing3}

\renewcommand\arraystretch{1.5}
\begin{center}
\begin{tabular}{|c|ccc|c|c|}
	\hline
	 & Cloud & Edge & Opportunistic Edge & Objective & Technique and solution\\
	\hline
	\multirow{2}{*}{G. Sum \emph{et al.} \cite{EnergyTradingFog}} &\multirow{2}{*}{SG} & \multirow{2}{*}{AG} & \multirow{2}{*}{\ding{55}} & Max EV energy utilization & Convex optimization \\
	\cline{5-6} & & & & Max operator and EV revenue & Genetic algorithm \\
	\hline 
    N. Kumar \emph{et al.} \cite{cloudSGnew} & SG & \ding{55} & EVs & Max EV utility & Bayesian coalition game \\
    	\hline
    M. Hossein \emph{et al.} \cite{fog1} & SG & Local energy market & home gateways & Min user and operator cost & Convex optimization \\
    \hline     
     D. A. Chekired \emph{et al.} \cite{FogEV} & SG & Microgrid & EVs & Max utility of grid, microgrid, and EVs & Linear programming \\
     \hline
     
\end{tabular}
\end{center}
\end{table*}

\subsubsection{Computing for V2G Scheduling}

Different from EV charging scheduling, V2G scheduling plays an important role in achieving the demand-supply balance at both system and user level. Therefore, computing of V2G scheduling needs to be fast and effective, and a hierarchical computing architecture can help achieve the scheduling performance. In this section, we review related works on computing for V2G scheduling in terms of their computing structure, as summarized in Table \ref{computing3}.

\begin{itemize}

\item \textbf{Two-tier computing} - With the SG operator at the upper tier, while deploying computing devices near users (\emph{e.g.,} grid-connected facilities), the two-tier computing hierarchy can effectively reduce energy/information exchange at the SG level. \cite{EnergyTradingFog} deploys local computing nodes at AGs to perform as energy/information interface among vehicles. Both non-profit and profit-driven scheduling processes are discussed and efficiently solved in \cite{EnergyTradingFog}. When the computing tasks are distributed further at EVs as in \cite{cloudSGnew}, EVs can cooperate to perform computation tasks and disseminate information to nearby EVs. The approach of using EVs as edge nodes can effectively enhance the information process performance and mitigate energy shortage. 

\item \textbf{Three-tier computing} - A more structured computing architecture considers cloud computing at the grid level, providing high computation capacity for day-ahead energy scheduling; AGs as edge nodes to provide a platform for customers to share and monitor data; and EVs as opportunistic edge nodes to perform real-time charging/discharging. In \cite{fog1}, a transactive energy management system is formed for energy exchange among end users. The work analyzes and validates the positive effect of multi-tier communication on bandwidth and delay performance. In \cite{FogEV}, a decentralized EV scheduling is achieved using hierarchical computing and SDN control. To reduce the peak load, a real-time dynamic pricing model is introduced for EV charging/discharging based on real-time demand-supply curve. It is concluded that offloading scheduling tasks to different tiers improves the grid stability and reduces response time of the tasks. 

\end{itemize}

\subsubsection{Security of V2G Technology}

Similar to EV charging, V2G demands secure payment protection. Due to the nature of V2G service (\emph{e.g.}, regulation and peak load mitigation), V2G payment needs to be fast, anonymous, and secure \cite{V2Gsecurity}. From the EV's perspective, its identity and location data should be unlinkable to AGs; service-related information (\emph{e.g.,} battery status, service preference, etc.) should be protected from adversaries. From the operators' perspective, they need to protect users' privacy and acquire appropriate data for efficient operation. \cite{V2Gsecurity} presents a novel network security architecture for secure V2G payment. The trade-off in terms of the operator is studied in \cite{V2Gblockchain}, where the payment mechanism is a registration and data maintenance process based on blockchain. With the proposed scheme, the anonymity of user payment data and payment auditing can be achieved.

Authentication is crucial for V2G service to prevent malicious attacks and adversaries. \cite{INFOCOM} discusses the scenario where AGs are third-party entities and EV privacy could be disclosed during operation. A robust privacy-preserving authentication scheme is then proposed that utilizes variant pseudonyms of EVs to prevent location privacy leakage. \cite{lightassama} considers the case when EVs are malicious and proposes a lightweight connection scheme for V2G service. By generating pseudonym identities, EVs preserve their private information while the proposed scheme forces EVs to follow a specific procedure to prevent malicious attacks. \cite{V2Gmag} discusses secure V2G service processes in different contexts (\emph{i.e.}, battery statuses and EV roles) and proposes a context-aware authentication solution for V2G communication correspondingly.

\section{OPEN ISSUES}

The technical development of power, communication, and computing technologies pushes forward the development of the EVN, while also bring new challenging and research direction in the inter-disciplinary area. In this section, we discuss open issues at different EVN development stage.

\subsection{EV AG Deployment}

\begin{itemize}

\item \textbf{\emph{Sectionalize Traffic Pattern Model}:} As introduced in Section III, a variety of traffic pattern models have been proposed to model the vehicular mobility in the areas of communication and computation. Compared to generalize the vehicular traffic pattern using both statistic and stochastic methods, a more efficient modelling method is to perform vehicular mobility analysis by geographic section. In terms of different section functionality (\emph{e.g.}, freeway, residential, commercial, or industrial sections), the vehicular driving pattern also changes both spatially and temporally. By grasping the user behaviour patterns in different sections, the EV traffic pattern can be effectively modelled.

\item \textbf{\emph{V2G Enabled AG Deployment:}} The capability of feeding energy back makes EVs unique components in the SG. The promising V2G technology is an essential part of the EVN that should be considered starting from the AG deployment stage. In this case, the AG needs to estimate its potential V2G service capacity, when taking factors of infrastructure size and V2G service profits into account. Moreover, the impact of reverse power flow of EV discharging to the SG also needs to be well studied to alleviate the EV integration influence on the system.

\item \textbf{\emph{SDN-based EVN Operation:}} While SDN presents to be a flexible, effective, and cost-efficient control method for the EVN, it still has many open issues that need further investigation. First, the centralized control manner not only makes SDN vulnerable to the single point of failure \cite{SDNsurvey2019}, but also reduces the EVN resilience. One solution to this issue is through decentralized SDN operation, in which case, communication overhead, census, and synchronization problems need to be addressed. Further, because of the unique features of the EVN (\emph{e.g.}, energy flow, dynamic topology, heterogeneous communication environment, etc.), the implementation of SDN is a fundamental problem. While there have been test-beds developed for SDN and SG, respectively\cite{SDNsurvey2019}, a test-bed integrating both SDN and EVN requires further study.

\end{itemize}

\subsection{EV Charging Scheduling}

\begin{itemize}

\item \textbf{\emph{Charging on-the-move}:} Empowered by the advanced inductive power transfer technique, EVs are expected to be charged on-the-move, and therefore effectively alleviate their range anxiety issues \cite{Info5}. Wireless charging system consists of primary coils deployed on the roadway system and secondary coils on-board to pick up the transferred energy. While the system has been proved experimentally feasible by several research projects \cite{koreaWPT}, technical issues such as coil structure design, power supply scheme, and segmentation switching techniques still require development. Moreover, by relaxing the range anxiety issues from EV charging, the charging scheduling becomes more flexible, which provides the AG operator with potentials to enhance operation performance.

\item \textbf{\emph{Trajectory Modeling of SAG Integration:}} Performing the resource allocation in the multi-dimensional heterogeneous SAG environment requires the modelling of different trajectories of satellites and HAP/LAP for efficient resource allocation. Without proper movement modelling, the SAG system could encounter frequent handover. The mobility model of mobile ad hoc networks (MANETs), random waypoint could be a potential modelling method \cite{sagsurvey190}. However, the high-speed node movements and network interference from other segment could still impede a smooth data communication.

\item \textbf{\emph{Information Security and Privacy:}} While security issues in vehicular communication have been well-studied, the unique feature of EVN operation poses new challenges from different perspectives. When scheduling EV charging, secure payment is demanded to establish mutual authentication in a fast, efficient, and traceable manner. The trade-off between vehicular privacy and security of payment needs to be carefully studied. In addition to considering malicious attacks in the SG, EV itself encounters more challenges to address malicious attacks from different entities (\emph{e.g.,} SG, AG, nearby EVs), as summarized in \cite{IoEV}. Attack detection and risk assessment are crucial for EVN operation security and can be a promising research topic.

\end{itemize}

\subsection{V2G Technology}

\begin{itemize}

\item \textbf{\emph{The Implementation of V2G:}} As the traditional power grid considers mostly uni-direction power transmission from the generation system to the loads, the existing transmission infrastructure cannot meet the technical demand of V2G technology. To achieve the bidirectional power flow, power electronic interfaces between EVs and the SG need to be well-developed for timely information update and demand delivery \cite{electron,infrasr}. The bidirectional power electronics need AC/DC converters to correct the power factor along with DC/DC converters to regulate the battery current within a reasonable range \cite{infrasr}. Such a complicated circuit topology requires necessary safety measurement and distribution system upgrade to guarantee the robustness of V2G implementation \cite{infrasr}. 

\item \textbf{\emph{Mobile Energy Storage}:} The EV mobility is a double-edged sword to the SG: on one hand, it poses challenges to charging demand/V2G service modelling. On the other hand, it invokes the EV potentials as energy storage devices on-the-move. When being distributed properly, EVs can efficiently mitigate the SG energy imbalance condition by delivering energy from energy-dense areas to energy-sparse areas. However, delivering energy with a large fleet of EV has the potential to cause unexpected traffic congestion, coupling the operation between the SG and the TN, which is an interesting research issue \cite{nannetwork}.

\item \textbf{\emph{EVN Service Provision through Network Slicing:}} As V2G services are provided to different electricity sub-markets, the energy and information resource needs to be allocated so as to meet requirements of different service. To exploit the full potential of network resources and optimize system performance, network slicing is adopted to allocate resources to particular users for their specific service requirements based on network virtualization technique \cite{NetworkSlicing, AIProceeding}. The technique has been applied in the next-generation network, and can be applied in the EVN by extending the resource allocation on power resource. However, the extension does require the implementation of power routers \cite{nansdn} while slicing challenges such as service operation specification still need further development.

\item \textbf{\emph{Information Security and Privacy:}} Compared to EV charging, information security and privacy issue is even more crucial in V2G scenario. For example, by jamming the communication channel, the V2G service requests between AGs and EVs could be interrupted, resulting in service demand and even energy imbalance at the system level. Further, the vehicular privacy such as its identity, location, battery usage information needs to be preserved while enough information is provided to AGs for scheduling V2G service. Therefore, security and privacy preservation scheme needs to be designed to addressed potential security threats in the cyber-physical system of EVN.

\end{itemize}

\section{CONCLUSIONS}

In this paper, we have presented a comprehensive survey on the EVN management from both energy and information perspectives. In particular, we have introduced the management framework incorporating SAG-integrated vehicular network, computing infrastructure, and EV-integrated SG. Research works on the AG deployment have been reviewed from energy, communication, and computation aspect to lay a solid basis for EVN management. EV scheduling works have been surveyed considering power flow analysis, communication requirements, and computation efficiency. To facilitate the EVN development, there are still challenges ahead. The lack of EV data under different traffic scenarios makes the EV traffic pattern estimation a challenging problem. Moreover, the reversing power flow incurred by V2G requires intensive research on the power system analysis and power infrastructure upgrade. Meanwhile, the emerging communication technologies such as SAG and 5G also demand proper resource allocation and deployment for timely data transfer.
This survey paper can shed the lights on the research of EVN considering inter-disciplinary management of energy and information. To facilitate smooth operation of EVN, the outlined open issues should be further investigated.

\bibliographystyle{IEEEtran}
\bibliography{buffer}

	\vspace*{-2.0\baselineskip}
	\small{
	\begin{IEEEbiography}
	[{\includegraphics[width=1in,height=1.25in,clip,keepaspectratio]{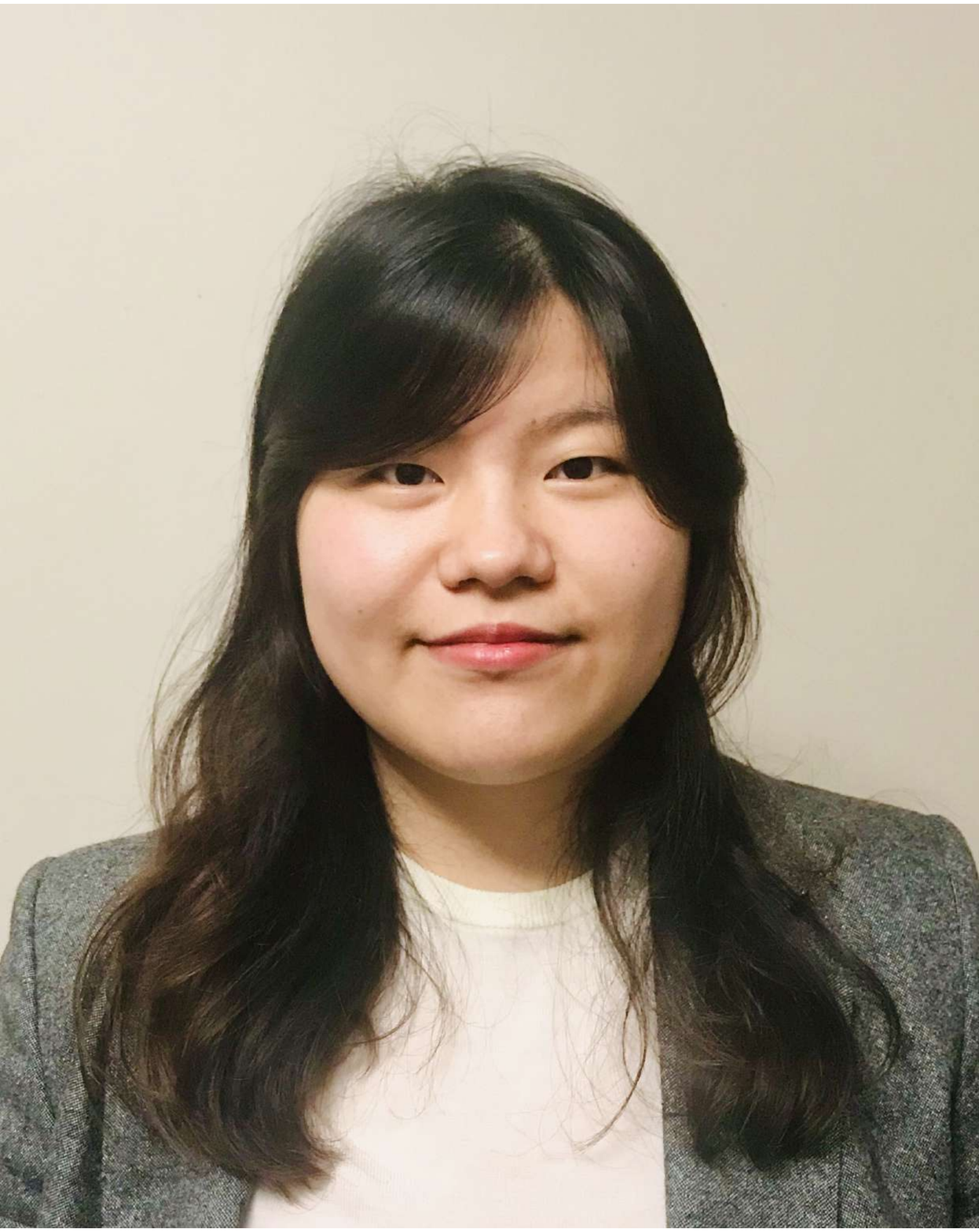}}]{Nan Chen}(M'20) received her Bachelor degree in electrical engineering and automation from Nanjing University of Aeronautics and Astronautics, Nanjing, Jiangsu, China, in 2014, and the Ph.D. degree in electrical and computer engineering from the University of Waterloo, Waterloo, ON, Canada, in 2019. Her current research interests include electric vehicle charging/discharging scheme design in smart grid, next-generation wireless networks, and machine learning application in vehicular cyber-physical systems.
	\end{IEEEbiography}}

	\vspace*{-2.0\baselineskip}
	\small{
	\begin{IEEEbiography}[{\includegraphics[width=1in,height=1.25in,clip,keepaspectratio]{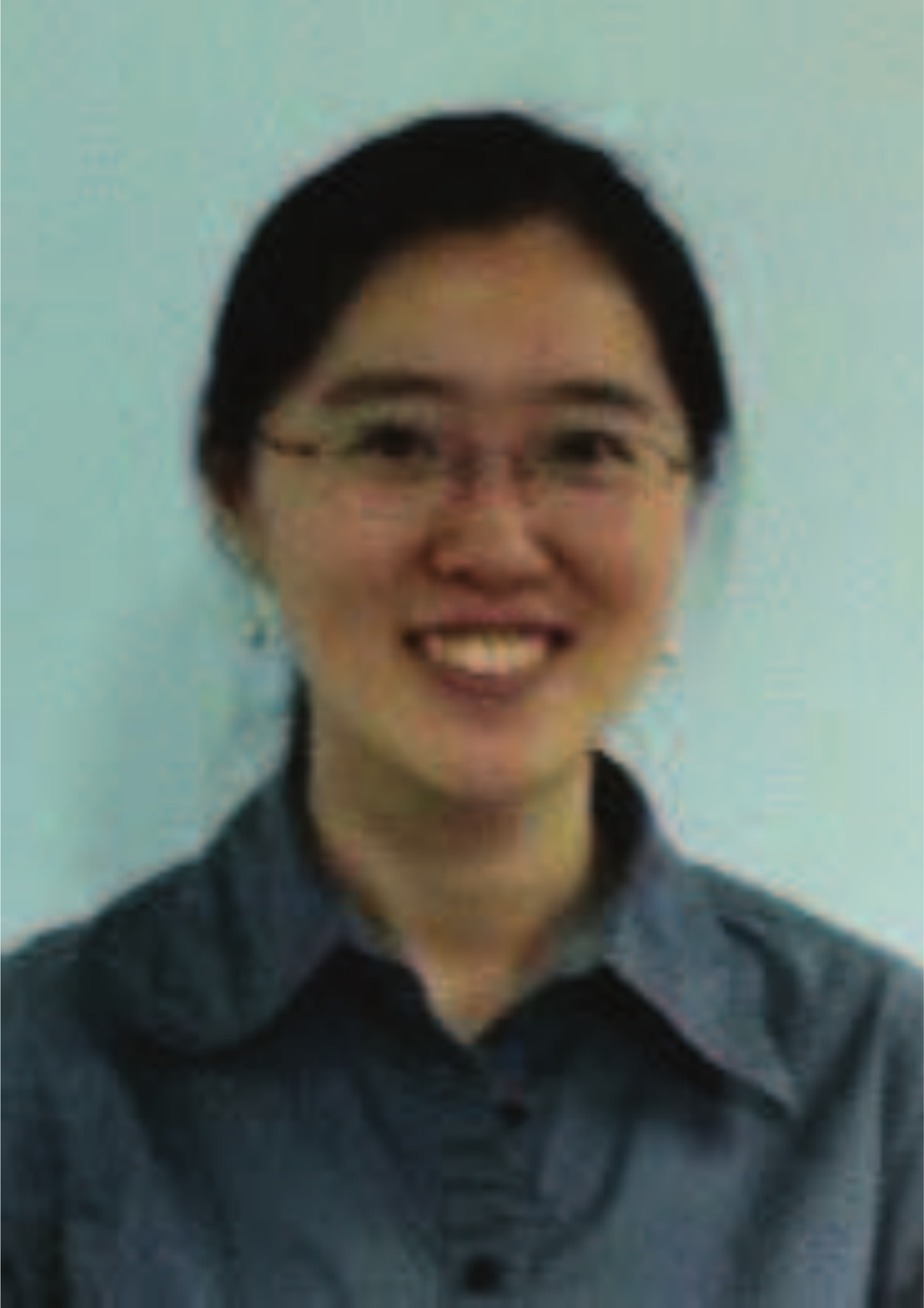}}]{Miao Wang}(M'15) is an Assistant Professor with the Department of Electrical and Computer Engineering, Miami University, Ohio, USA. She received her B.Sc. degree from Beijing University of Posts and Telecommunications and M.Sc. degree from Beihang University, Beijing, China, in 2007 and 2010, respectively, and the Ph.D. degree in electrical and computer engineering from the University of Waterloo, Waterloo, ON, Canada, in 2015. Her current research interests include electric vehicles charging/discharging strategy design in smart grid, traffic control, capacity and delay analysis, and routing protocol design for vehicular networks.
	\end{IEEEbiography}}
	
	\vspace*{-2.0\baselineskip}
	\small{
	\begin{IEEEbiography}[{\includegraphics[width=1in,height=1.25in,clip,keepaspectratio]{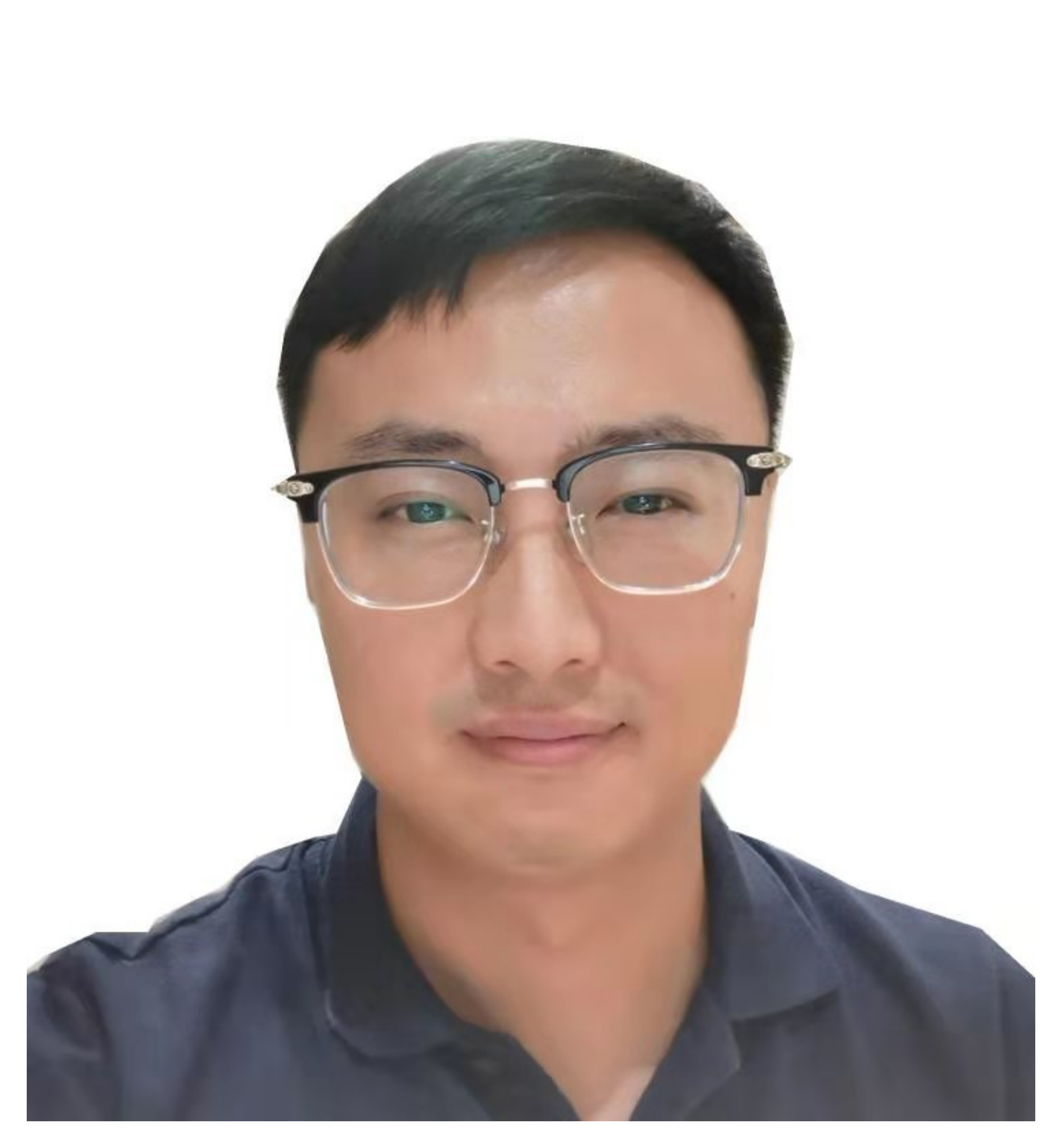}}]{Ning Zhang}(M'15-SM'18) received the Ph.D degree from University of Waterloo, Canada, in 2015. Between 2015 and 2017, he was a postdoc research fellow at University of Waterloo and University of Toronto, Canada, respectively. His research interests include wireless networking, mobile edge computing, and network security. So far, he has published over 130 articles, including 90+ IEEE journals and 40+ conference papers. He serves as an Associate Editor of IEEE Internet of Things Journal, IEEE Transactions on Cognitive Communications and Networking, IEEE Access and IET Communications, and an Area Editor of Encyclopedia of Wireless Networks (Springer). He also serves/served as a Guest Editor for several international journals, such as IEEE Wireless Communications and IEEE Transactions on Cognitive Communications and Networking, and IEEE Transactions on Industrial Informatics. He serves/served as the track chair for several international conferences, such as IEEE VTC and EAI AICON, and a co-chair for several international workshops. He received the Best Paper Awards from IEEE Globecom in 2014, IEEE WCSP in 2015, and Journal of Communications and Information Networks in 2018, IEEE ICC in 2019, IEEE Technical Committee on Transmission Access and Optical Systems in 2019, and IEEE ICCC in 2019, respectively. He has been a senior member of IEEE since 2018.
	\end{IEEEbiography}}

	\vspace*{-2.0\baselineskip}
	\small{
	\begin{IEEEbiography}[{\includegraphics[width=1in,height=1.25in,clip,keepaspectratio]{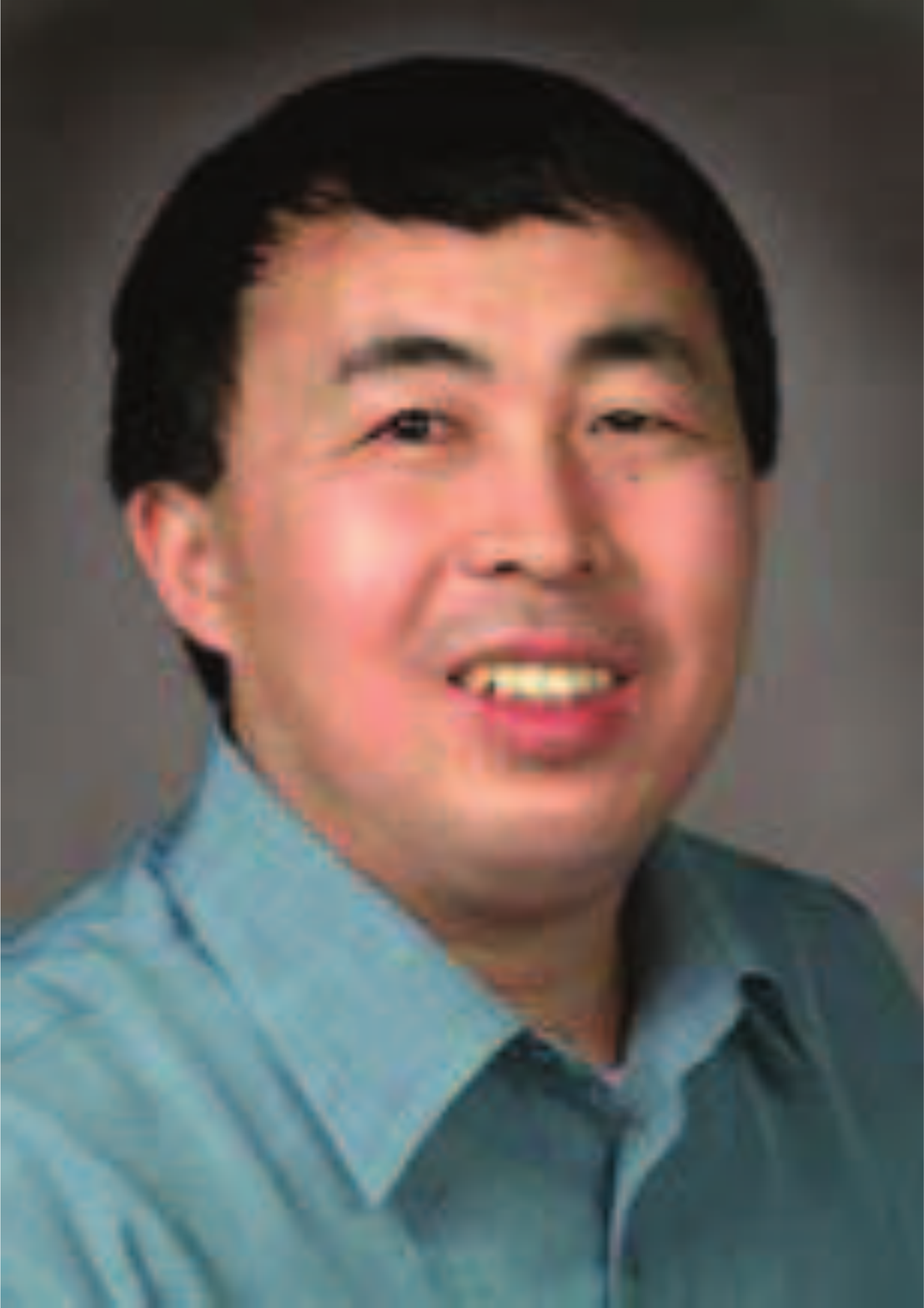}}]{Xuemin (Sherman) Shen} (M'97-SM'02-F'09) received the B.Sc. degree from Dalian Maritime University, Dalian, China, in 1982, and the M.Sc. and Ph.D. degrees from Rutgers University, Camden, NJ, USA, in 1987 and 1990, respectively, all in electrical engineering. He is a University
Professor and the Associate Chair for Graduate Studies with the Department of Electrical and
Computer Engineering, University of Waterloo, Waterloo, ON, Canada. His research interests include resource management, wireless network security, social networks, smart grid, and vehicular ad hoc and sensor networks. He was the Technical Program Committee Chair/Co-Chair for IEEE Globecom‘16, Infocom’14, IEEE VTC‘10 Fall, and Globecom’07, the Symposia Chair for IEEE ICC’10, the Tutorial Chair for IEEE VTC’11 Spring, and IEEE ICC’08, the General Co-Chair for ACM Mobihoc’15, Chinacom’07,
and the Chair for IEEE Communications Society Technical Committee on Wireless Communications. He is/was also the Editor-in-Chief for the IEEE Internet of Things Journal, IEEE Network, Peer-to-Peer Networking and Application, and IET Communications, a Founding Area Editor for the IEEE Transactions on Wireless Communications, an Associate Editor for the IEEE Transactions on Vehicular Technology, Computer Networks, and ACM/Wireless Networks, etc., and a Guest Editor for the IEEE Journal on Selected Areas in Communications, IEEE Wireless Communications, and IEEE Communications Magazine. He was the recipient of the Excellent Graduate Supervision Award in 2006 and the Premiers Research Excellence Award (PREA) in 2003 from the Province of Ontario, Canada. He is a registered Professional Engineer of Ontario, Canada, an
Engineering Institute of Canada Fellow, a Canadian Academy of Engineering Fellow, a Royal Society of Canada Fellow, and a Distinguished Lecturer of IEEE Vehicular Technology Society and Communications Society.

\end{IEEEbiography}}

\end{document}